\documentclass[twocolumn]{aastex61}

\usepackage{graphicx}
\usepackage{natbib}
\usepackage{amsmath}
\usepackage{amssymb}
\usepackage{epstopdf}
\usepackage{xcolor}

\begin{document}

\title{HST Scattered Light Imaging and Modeling of the Edge-on Protoplanetary Disk ESO-H$\alpha$ 569}
%\author{Schuyler Wolff}

\shorttitle{ESO H$\alpha$ 569 Radiative Tranfer Modeling}
\shortauthors{Wolff et al.}

\correspondingauthor{Schuyler G. Wolff}
\email{wolff@strw.leidenuniv.nl}

\author{Schuyler G. Wolff}
\affil{Johns Hopkins University \\ Baltimore, MD 21218, USA}
\affil{Leiden Observatory, Leiden University \\ 2300 RA Leiden, The Netherlands}

\author{Marshall D. Perrin}
\affil{Space Telescope Science Institute \\ Baltimore, MD 21218, USA}

\author{Karl Stapelfeldt}
\affil{Exoplanet and Stellar Astrophysics Laboratory \\ NASA Goddard Space Flight Center}
\affil{Jet Propulsion Laboratory \\ California Institute of Technology}

\author{Gaspard Duch\^{e}ne}
\affil{University of California, Berkeley}
\affil{Universit\'e Grenoble-Alpes, CNRS \\ Institut de Plan\'etologie et d'Astrophyisque (IPAG) \\ F-38000 Grenoble, France}

\author{Francois M\'{e}nard}
\affil{Universit\'e Grenoble-Alpes, CNRS \\ Institut de Plan\'etologie et d'Astrophyisque (IPAG) \\ F-38000 Grenoble, France}

\author{Deborah Padgett}
\affil{Jet Propulsion Laboratory \\ California Institute of Technology}

\author{Christophe Pinte}
\affil{UMI-FCA, CNRS/INSU, France (UMI 3386)}
\affil{Dept. de Astronom\'{\i}a \\ Universidad de Chile \\ Santiago, Chile}

\author{Laurent Pueyo}
\affil{Space Telescope Science Institute \\ Baltimore, MD 21218, USA}

\author{William J. Fischer}
\affil{Space Telescope Science Institute \\ Baltimore, MD 21218, USA}

\begin{abstract}
We present new HST ACS observations and detailed models for a recently discovered edge-on protoplanetary disk around ESO~H$\alpha$~569 (a low-mass T Tauri star in the Cha~I star forming region). Using radiative transfer models we probe the distribution of the grains and overall shape of the disk (inclination, scale height, dust mass, flaring exponent and surface/volume density exponent) by model fitting to multiwavelength (F606W and F814W) HST observations together with a literature compiled spectral energy distribution. A new tool set was developed for finding optimal fits of MCFOST radiative transfer models using the MCMC code \texttt{emcee} to efficiently explore the high dimensional parameter space. 
It is able to self-consistently and simultaneously fit a wide variety of observables in order to place constraints on the physical properties of a given disk, while also rigorously assessing the uncertainties in those derived properties.
We confirm that ESO~H$\alpha$~569 is an optically thick nearly edge-on protoplanetary disk. The shape of the disk is well described by a flared disk model with an exponentially tapered outer edge, consistent with models previously advocated on theoretical grounds and supported by millimeter interferometry. The scattered light images and spectral energy distribution are best fit by an unusually high total disk mass (gas+dust assuming a ratio of 100:1) with a disk-to-star mass ratio of 0.16. 
\end{abstract}


\section{Introduction}
\label{Sec:intro}

We seek to understand the initial conditions for planet formation and the physical processes that contribute to the assembly of planets by measuring the properties of young protoplanetary disks. The unique geometry of edge-on circumstellar disks provides a valuable opportunity to study detailed disk structure, as the bright central star is occulted from view and thus does not pose a contrast problem. The width of the disk's dark lane (the vertical extent of the $\tau=1$ surface), outer radius, and degree of flaring can be directly measured, and the scale height of the disk can be related to the local disk temperature \citep{2007prpl.conf..523W}.
\citet{2004IAUS..202..291S}  provides a review of the observational advantages of targeting edge-on disks. 
%and grain distributions in the disk. 
Previous studies of edge-on disks have measured disk inclinations and dust masses from a combination of scattered light images and millimeter continuum maps
\citep{2003ApJ...588..373W,2009A&A...505.1167S}. Additionally, the change in the dust lane thickness with wavelength allows dust grain properties to be derived \citep{2001ApJ...556..958C, 2008ApJ...688.1118W, 2010ApJ...712..112D, 2011ApJ...727...90M}. However, the sample of edge-on disks with high resolution observations remains relatively small. 

ESO~H$\alpha$~569, a young M2.5 star embedded in the Chameleon I star forming region (SFR),
was imaged as part of an HST observation program designed to double the sample of edge-on protoplanetary disks for which high resolution scattered light images have been obtained. The sample for the survey was chosen from WISE and Spitzer surveys of nearby star forming regions (SFRs) which allow identification of new candidate edge-on disks from their characteristic double peaked spectral shape. HST program 12514 in Cycle 19 obtained high resolution optical imaging of the top 21 candidates, including the data presented in the current work \citep{2014IAUS..299...99S}.  
%Several of the targets in this sample, including ESO~H$\alpha$~569, are known members of the Chameleon star forming region (SFR). 

Several of the targets in this sample of edge-on protoplanetary disks, including ESO~H$\alpha$~569, are known members of the Chameleon I (Cha I) SFR. 
Distances to Cha I, one of the nearest SFRs, have been determined in a variety of ways including zero-age main sequence fitting and Hipparcos parallaxes of members. \citet{1997A&A...327.1194W} provide a review of the results and combine measurements to arrive at a distance of $160 \pm 15$ pc. \citet{1999A&A...352..574B} confirm this distance after cross-correlating the Herbig \& Bell and Hipparcos Catalogues. 
\citet[][see Appendix B1]{2011A&A...527A.145B} present a more detailed review of Cha I distance measurements.
Age estimates for Chamaeleon I range from 1 - 2 Myrs \citep{1998A&A...337..403B,2000ApJ...542..464C}. 
The Cha I SFR is characterized by a relatively high extinction with an observed maximum of $A_{V} \sim 10$ \citep{1997A&A...324L...5C}. Such a high extinction would suppress the blue side of the spectral energy distribution of a young stellar system. 
The initial mass function for Cha I has a maximum mass of $0.1 - 0.15 \, M_{\odot}$  \citep{2007ApJS..173..104L}, while the total mass of Cha I is $\sim$ 1000 $M_{\odot}$ \citep{1998A&A...332..273B}.

The remainder of this introduction summarizes prior observations of ESO~H$\alpha$~569. 
In Section \ref{Sec:obs}, we present high resolution HST scattered light observations of the ESO~H$\alpha$~569 protoplanetary disk and a spectral energy distribution (SED) compiled from the literature. In Section \ref{Sec:model}, radiative transfer modeling efforts to fit these observations to a variety of disk properties are discussed. Both a grid and Monte Carlo Markov Chain (MCMC) approach were used to explore parameter space, and results are given in Sections \ref{Sec:gridres} and \ref{Sec:mcmcres}.
Section \ref{Sec:discussion} discusses these results including the gravitational stability of the system and places ESO~H$\alpha$~569 in context with previous disk observations. Lastly, Section \ref{Sec:summary} provides a Summary and Conclusions.

\subsection{Prior Studies of ESO~H$\alpha$~569}

ESO~H$\alpha$~569 (2MASS J11111083-7641574) 
was first identified as a target of interest in the \citet{2004A&A...417..583C} European Southern Observatory survey of young stars with strong H$\alpha$ emission in Cha I SFR.  
\citet{2004A&A...417..583C} classified the central star as K7 using ground-based spectroscopy.
The authors noted that this object is severely under-luminous for a K7 star (by $\sim$ 2 orders of magnitude), which made it a prime candidate for our edge-on disks survey. 
The \citet{2007ApJS..173..104L} survey of the stellar population in Chamaeleon obtained an R$\approx$5000 spectrum from 0.6-0.9 $\mu m$, which gave a spectral type of M2.5, an effective temperature of 3488 K, and an apparent bolometric luminosity of $L_{bol}$ =  0.0030 $L_{\odot}$. More recently,  broad-band spectroscopy with VLT/X-Shooter provides a spectral type of M1 $\pm$ 2 subtypes \citep[][their table 3]{2017arXiv170402842M}, and confirms that the target appears underluminous. Because ESO~H$\alpha$~569 is heavily extincted by the disk, the apparent luminosity is an unreliable estimator for the true bolometric luminosity of the central star. For stars of the same spectral type in the \citet{2007ApJS..173..104L} survey, the average bolometric luminosity is $0.34 \pm 0.08 L_{\odot}$. The current study adopts this luminosity and a $\sim$ 3500 K effective temperature. Using the theoretical evolutionary models of \citet{1998A&A...337..403B} for low mass stars with solar metallicity gives a mass for the central star of 0.35 $M_{\odot}$. The associated stellar radius is 1.13 $R_{\odot}$.

Prior attempts have been made to infer the disk properties of ESO H$\alpha$ 569 based on its spectral energy distribution. \citet{2007ApJS..173..104L} noted that the X-ray non-detection of this star indicates an extinction of $A_K \geq 60$, consistent with obscuration by an edge-on disk, assuming its X-ray luminosity is that of a typical T Tauri star.
\citet{2012AJ....144...83R} combine published 2MASS and Spitzer photometry, with unresolved HST fluxes to fit properties of the disk and central star using the online library of 20,000 models of young circumstellar systems compiled by \citet{2006ApJS..167..256R}. These models include the central star, a diffuse envelope and an accreting disk \citep{2003ApJ...598.1079W, 2003ApJ...591.1049W, 2004ApJ...617.1177W}.  The authors find the disk is best fit by an inclination of $\sim$ 87.1 degrees, $L_{bol} = 0.8 \pm 0.4 L_{\odot}$, $M_{star} = 0.33 \pm 0.03 M_{\odot}$,  $R_{\mathrm{star}} = 2.5 \pm 0.6 R_{\odot}$, and give an upper limit for the sub-mm disk mass of $0.005 M_{\odot}$.  \citet{2014MNRAS.443.1587R} included Herschel data and found a best fit inclination of 81.4 degrees.
More recently, \citet{2016ApJ...831..125P} provide 1.3 millimeter continuum data which corresponds to a disk mass estimate of $0.0046 \, M_{\odot}$, using an assumed opacity of $\kappa = 2.3 cm^{2}/g$, a gas to dust ratio of 100, and a disk temperature of 20~K.

\begin{figure}[htpb!]
\begin{center}
\includegraphics[width=3.1in]{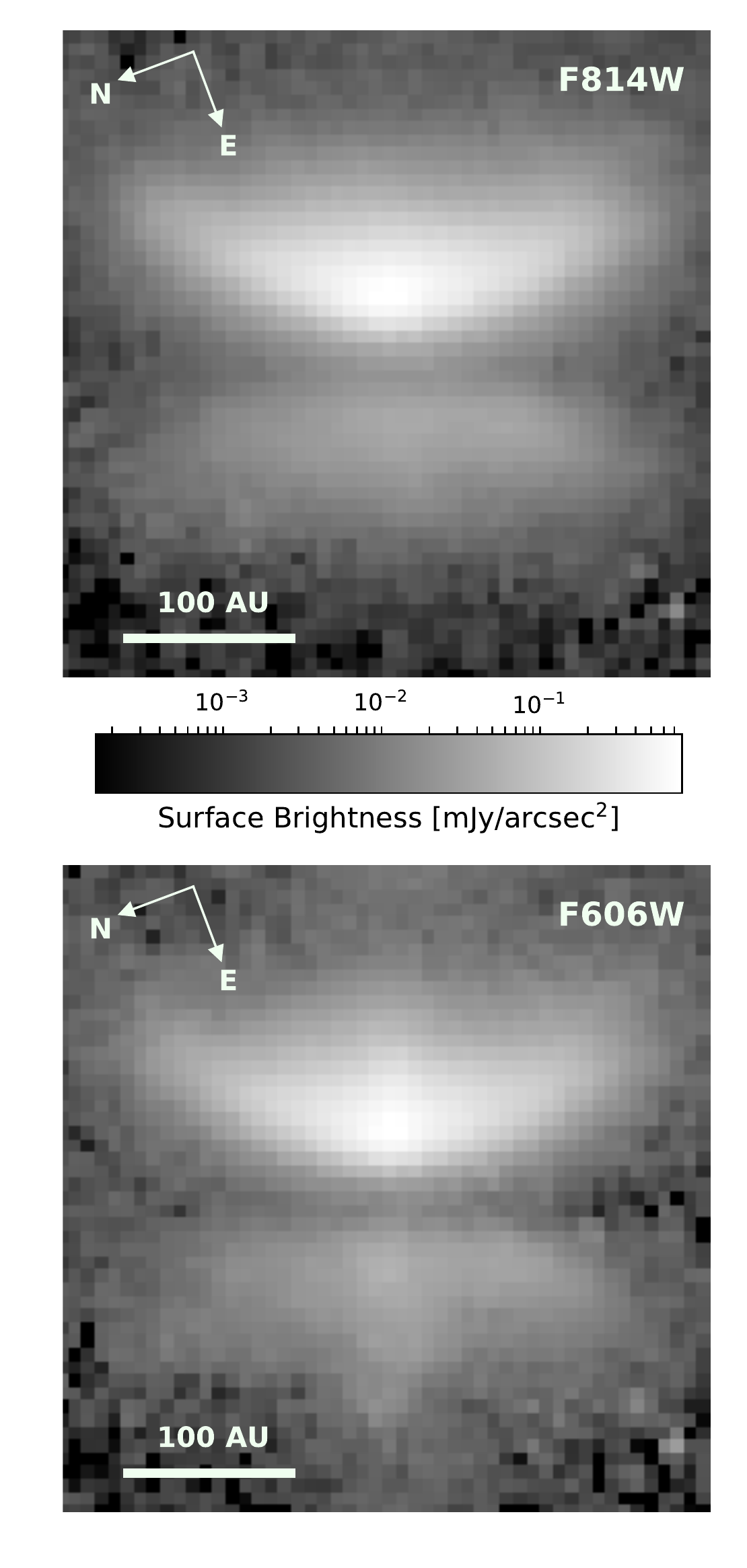}
\caption{HST images of the protoplanetary disk ESO-H$\alpha$569. Top: F814W. Bottom: F606W. Both images show the dark dust lane and asymmetries between the top and bottom of the disk, while only F606W establishes the presence of an outflow jet. The 100 au scale bar corresponds to an angular scale of 0.625$^{\prime\prime}$.} \label{images}
\end{center}
\end{figure}

\section{Observations}
\label{Sec:obs}

\subsection{HST Scattered Light Images}

%\startlongtable
\begin{deluxetable*}{ccccccc}[!t]  % <--- column justification (center/left/right)
\tablecolumns{7}
\tablecaption{Spectral energy distribution photometry and references.}
\tablehead{   % column headings
  \colhead{$\lambda (\mu m)$} &
  \colhead{Flux (mJy)} &
%  \colhead{Flux (mag)} &
  \colhead{Source} &
  \colhead{Instrument} &
  \colhead{Bandwidth ($\mu m$)} &
  \colhead{Angular Resolution} &
  \colhead{Date}
}
\tablewidth{6.9in}
%\colnumbers
\startdata
0.551 & 0.030 $\pm$ 0.004 & Robberto et al. 2012 & HST WFPC2 & 0.14 &  0.0996$^{\prime\prime}$ & 2009-04-27 \\
0.606 & 0.058 $\pm$ 0.001 & This work & HST ACS & 0.27 & 0.05$^{\prime\prime}$  &  2012-03-09 \\   %0.467 - 0.7349 
0.814 & 0.21 $\pm$ 0.01 & This work  & HST ACS & 0.31 & 0.05$^{\prime\prime}$  & 2012-03-09 \\   %0.6948 - 1.0043
1.235 & 0.66 $\pm$ 0.05 & Skrutskie et al. 2006 & 2MASS & 0.16 & $\sim$5$^{\prime\prime}$ & 2000-01-25  \\ %1.105 - 1.349
1.662 & 0.97 $\pm$ 0.08 & Skrutskie et al. 2006 & 2MASS & 0.25 & $\sim$5$^{\prime\prime}$ & 2000-01-25  \\ %1.504 - 1.709
2.15 & 0.98 $\pm$ 0.09 & Skrutskie et al. 2006 & 2MASS & 0.26 & $\sim$5$^{\prime\prime}$ & 2000-01-25 \\ %1.989 - 2.316
3.6 & 0.58 $\pm$ 0.03 & Luhman et al. 2008 & Spitzer IRAC & 0.75 & $\sim$2$^{\prime\prime}$ &  2004-07-04  \\ %3.179 - 3.955
4.5 & 0.57 $\pm$ 0.05 & Luhman et al. 2008 & Spitzer IRAC & 1.02 & $\sim$2$^{\prime\prime}$ & 2004-07-04   \\ %3.955 - 5.015
5.8 & 0.58 $\pm$ 0.05 & Luhman et al. 2008 & Spitzer IRAC & 1.43 & $\sim$2$^{\prime\prime}$ &  2004-07-04  \\  %5.015 - 6.442
8.0 & 0.67 $\pm$ 0.05 &  Luhman et al. 2008 & Spitzer IRAC & 2.91 & $\sim$2$^{\prime\prime}$ & 2004-07-04 \\ %6.442 - 9.343 
3.4 & 0.63 $\pm$ 0.02 & Cutri et al. 2012 & WISE & 0.66 &  6.1$^{\prime\prime}$ & 2010-02-13,20 \\
4.6 & 0.71 $\pm$ 0.02 & Cutri et al. 2012 & WISE & 1.04 & 6.4$^{\prime\prime}$ & 2010-02-13,20 \\
12 & 0.65 $\pm$ 0.07 & Cutri et al. 2012 & WISE & 5.51 & 6.5$^{\prime\prime}$ & 2010-02-13,20 \\
22 & 7.5 $\pm$ 0.89 & Cutri et al. 2012 & WISE & 4.10 & 12.0$^{\prime\prime}$ & 2010-02-13,20  \\
24 & 8.36 $\pm$ 0.77 & Luhman et al. 2008 & Spitzer MIPS & 5.3 & 6$^{\prime\prime}$ & 2004-04-08  \\ %20.335 - 29.462
70 & 107 $\pm$ 10.8  & Luhman et al. 2008 & Spitzer MIPS & 19 & 18$^{\prime\prime}$ & 2004-04-08 \\ %60.92681 - 80.58062
70 & 200 $\pm$ 100 & Winston et al. 2012 & Herschel PACS & 25 & 5.8$^{\prime\prime}$ & 2011-06-23  \\ %60 - 90 
160* & 200 $\pm$ 200 & Winston et al. 2012 & Herschel PACS & 85 & 12.0$^{\prime\prime}$ & 2011-06-23 \\ %130 - 210
250* & 100 $\pm$ 100 & Winston et al. 2012 & Herschel SPIRE & 25 & 18$^{\prime\prime}$ & 2011-06-23  \\ %60 - 90 
350* & 50 $\pm$ 50 & Winston et al. 2012 & Herschel SPIRE & 25 & 25$^{\prime\prime}$ & 2011-06-23  \\ %60 - 90 
500* & 50 $\pm$ 50 & Winston et al. 2012 & Herschel SPIRE & 25 & 37$^{\prime\prime}$ & 2011-06-23  \\ %60 - 90 
870 & 72 $\pm$ 14 & Belloche et al. 2011 & APEX/LABOCA & 150 & 19.2$^{\prime\prime}$ & May 2008  \\  %(313 - 372 GHz) 
2830 & 3.2 $\pm$ 0.1 & Dunham et al. 2016 & ALMA & 55 & $\sim$2$^{\prime\prime}$ &  2013-11-29 to \\
 & & & & & & 2014-03-08 \\
\enddata
\tablecomments{Photometry at wavelengths marked with an * represent only upper limits and are not included in the spectral energy distribution modeling. }
\label{table:sed}
\end{deluxetable*}
%\end{center}
% Filter widths from http://casa.colorado.edu/~ginsbura/filtersets.htm
% IRAC bandpasses: http://irsa.ipac.caltech.edu/data/SPITZER/docs/irac/iracinstrumenthandbook/7/
% MIPS bandpasses: http://irsa.ipac.caltech.edu/data/SPITZER/docs/mips/mipsinstrumenthandbook/3/
% PACS bandpasses: http://herschel.esac.esa.int/Docs/PACS/pdf/pacs_om.pdf
% HST bandpasses: http://svo2.cab.inta-csic.es/theory/fps/index.php?id=HST/WFPC2.f814w&&mode=browse&gname=HST&gname2=WFPC2
% PACS zero points: http://svo2.cab.inta-csic.es/theory/fps/index.php?id=Herschel/Pacs.blue&&mode=browse&gname=Herschel&gname2=Pacs
% Laboca bandpass: http://www.apex-telescope.org/bolometer/laboca/technical/
% ALMA bandpass is 2 Ghz.

Scattered light images of the ESO~H$\alpha$~569 disk were obtained using HST ACS/WFC in both the F814W and F606W broad band filters on March 9th, 2012 as part of program GO 12514. The total exposure times were 1440 s for F606W and 960 s for F814, with each filter's exposure split as two integrations for cosmic ray rejection. 
The reduced and calibrated data produced by the HST pipeline were retrieved from the Mikulski Archive for Space Telescopes (MAST).

Figure \ref{images} provides the reduced images, rotated to place the disk major axis horizontal. 
The bipolar appearance unequivocally demonstrates the edge-on nature of ESO~H$\alpha$~569. 
%, which was until now only suspected based on its SED and a very marginal detection in prior HST observations by Robberto et al. (2008).
The western side is much brighter than the eastern (by $\sim 20\times$ comparing their peak surface brightnesses) and, along with the curvature of the nebula, indicates this side is tilted slightly toward us.   
There is no sign of starlight directly peeking through as an unresolved point source.
The position angle of the disk's minor axis was evaluated to be 65 $\pm$ 1 degrees. This was computed as the position angle for which mirroring the image across the minor axis minimized the flux difference between the left and right sides.  The disk is close to left/right symmetric, though the southern side (right side as shown in Fig. 1) is very slightly brighter.

%It gets much brighter toward the center of the disk on that side, closest to the star, but there's no sign of starlight directly peeking through as a point source

The disk is very red (much brighter in F814W than F606W).  The flux density of the disk was measured in both filters using a 50 pixel aperture, which corresponds to a spatial scale of 2" x 2" and was chosen to encompass all disk flux with surface brightness $\geq 3 \sigma$ above the background noise. The measured flux density is 0.058 $\pm$ 0.001 mJy for F606W and 0.21 $\pm$ 0.01 mJy for F814W, which gives a color [F814W]-[F606W] = 1.4 AB magnitudes. %3.62 $\pm$ 0.18. 

The disk has an apparent outer radius of 0.80 $\pm$ 0.05$^{\prime\prime}$ which corresponds to 125 $\pm$ 8 au at a distance of 160 pc. Here, the outer radius is inferred as the offset at which the flux declines to less than 10 \% the peak value for the widest part of the disk. 

%There is a jet clearly present in the F606W data oriented perpendicular to the disk axis. 

 \begin{figure}[hbpt!]
\begin{center}
\includegraphics[width=3.3in]{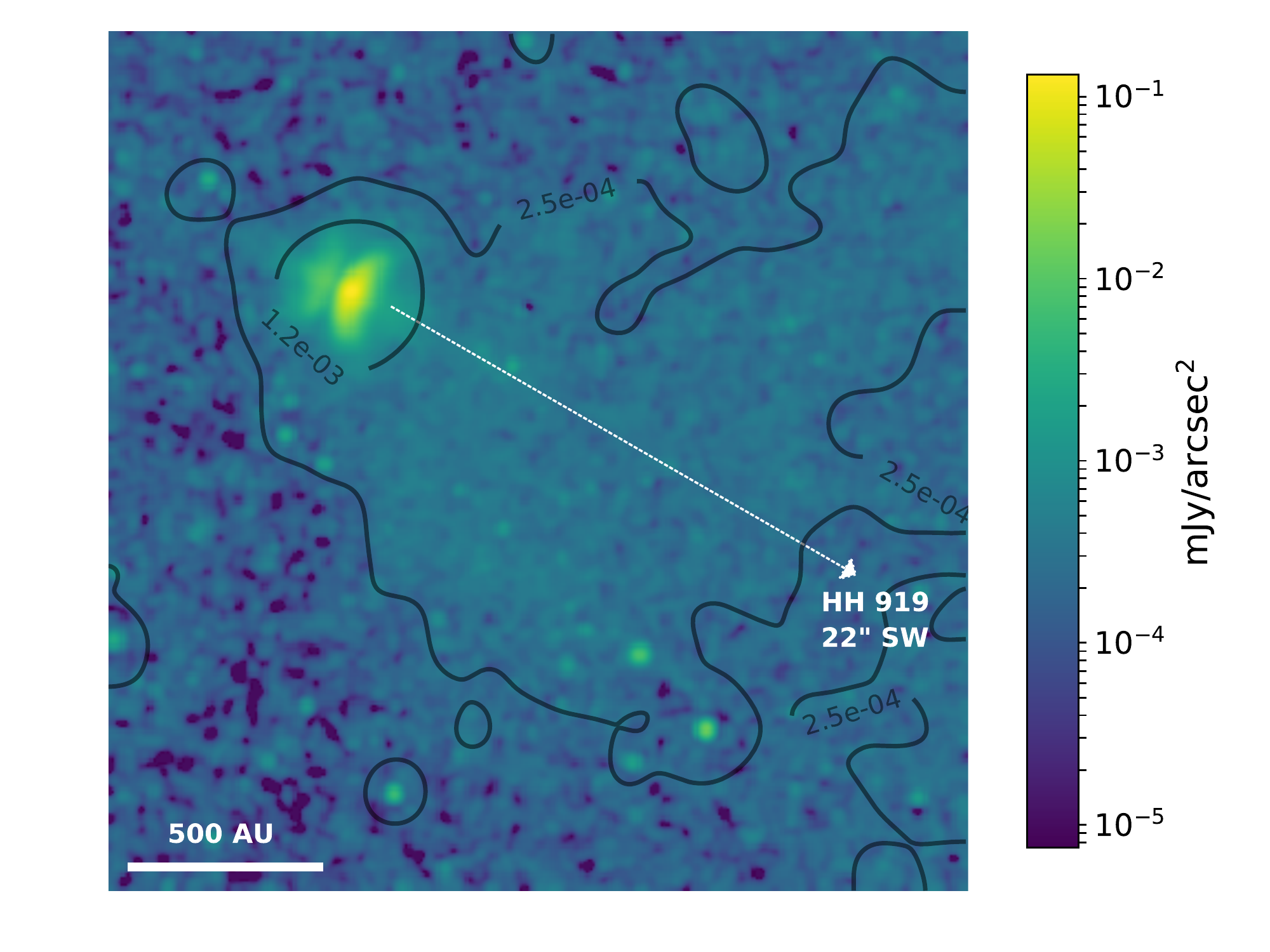}
\caption{A wider F606W filter image displaying the diffuse nebula extending outward from the disk. The direction of the H$\alpha$ filament HH919 is shown by the arrow. The jet lines up well with the reported position of HH 919, consistent with ESO~H$\alpha$~569 being the origin of this outflow. The 500 au scale bar corresponds to an angular scale of 3.125$^{\prime\prime}$. \label{zimage}}
\end{center}
\end{figure}

 \begin{figure}[hbpt!]
\begin{center}
\includegraphics[width=3.3in]{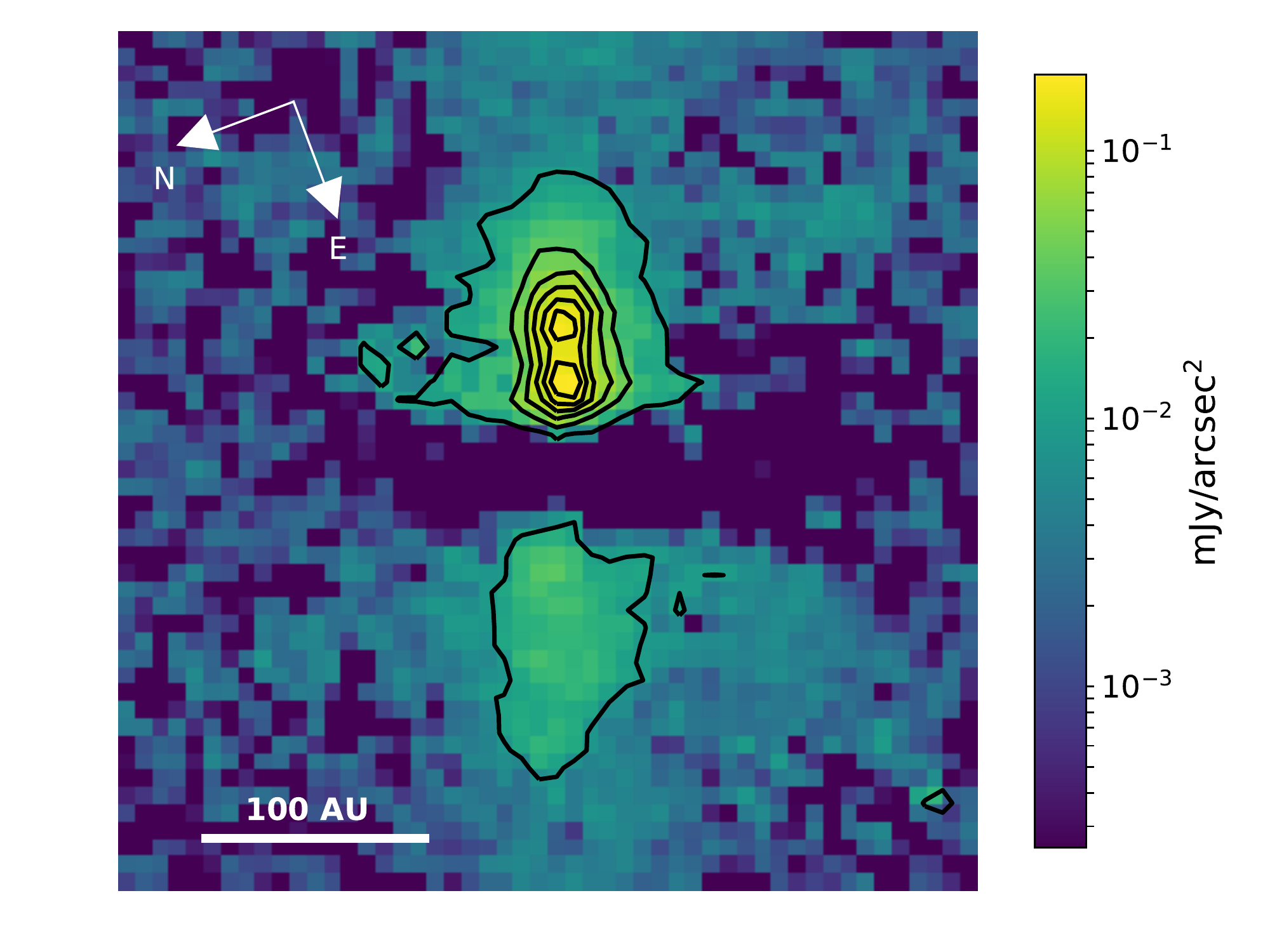}
\caption{An image of the jet created by subtracting the F814W image from the F606W image. Contours are drawn from 0.01 to 0.19 $\mathrm{mJy/arcsec}^{2}$ in intervals of 0.03 $\mathrm{mJy/arcsec}^{2}$. The 100 au scale bar corresponds to an angular scale of 0.625$^{\prime\prime}$.   \label{jetimage}}
\end{center}
\end{figure}

\subsection{HST Jet Outflow Images}

The strong H$\alpha$ emission in the spectrum of this young object indicates ongoing accretion onto the central star, which is often associated with launching of outflow jets. \citet{2006AJ....132.1923B} suggested ESO~H$\alpha$~569 as the possible source for the Herbig Haro object 919. HH 919 is an arcminute long filament with a PA of $\sim$ 60 - 75 degrees and is located 22$^{\prime\prime}$ (0.05 pc) southwest of ESO~H$\alpha$~569. 
A jet is visible in the F606W scattered light image extending vertically from the disk and is $\sim$ 0.25$^{\prime\prime}$ wide. 
This is consistent with the emission lines of H$\alpha$ and SII as are commonly seen in such outflows. 
%ESO~H$\alpha$~569 is located $\sim$ 22$^{\prime\prime}$ to the northeast of HH 919. 
A line connecting ESO~H$\alpha$~569 with HH 919 would have a PA of $\sim$ $63^{\circ}$, giving an orientation consistent with the ESO~H$\alpha$~569 jet serving as the culprit for the HH 919 filament.

Figure \ref{zimage} presents a wider field of view showing the interaction of this disk with the surrounding ISM. Diffuse nebulosity is visible extending outward from the disk. An image of the jet was created by subtracting the F814W image (scaled by a factor of 2.5) from the F606W image (Figure \ref{jetimage}).
The flux from the jet is difficult to decouple from the disk flux, but the jet accounts for roughly 50 \%  of the local surface brightness from the disk. This value is taken from an average of the flux over 9 pixels with the jet superimposed on the disk and compared to the flux in 9 neighboring pixels with no jet signature. The peak surface brightness of the jet is $\sim 0.19$~mJy/arcsec$^{2}$.
The ability to measure color variations in the shape of the disk and width of the dark lane between the F606W and F814W bands is hindered by the presence of this bright jet.

\subsection{Spectral Energy Distribution}

\begin{figure}[htpb!]
\begin{center}
\includegraphics[width=3.3in]{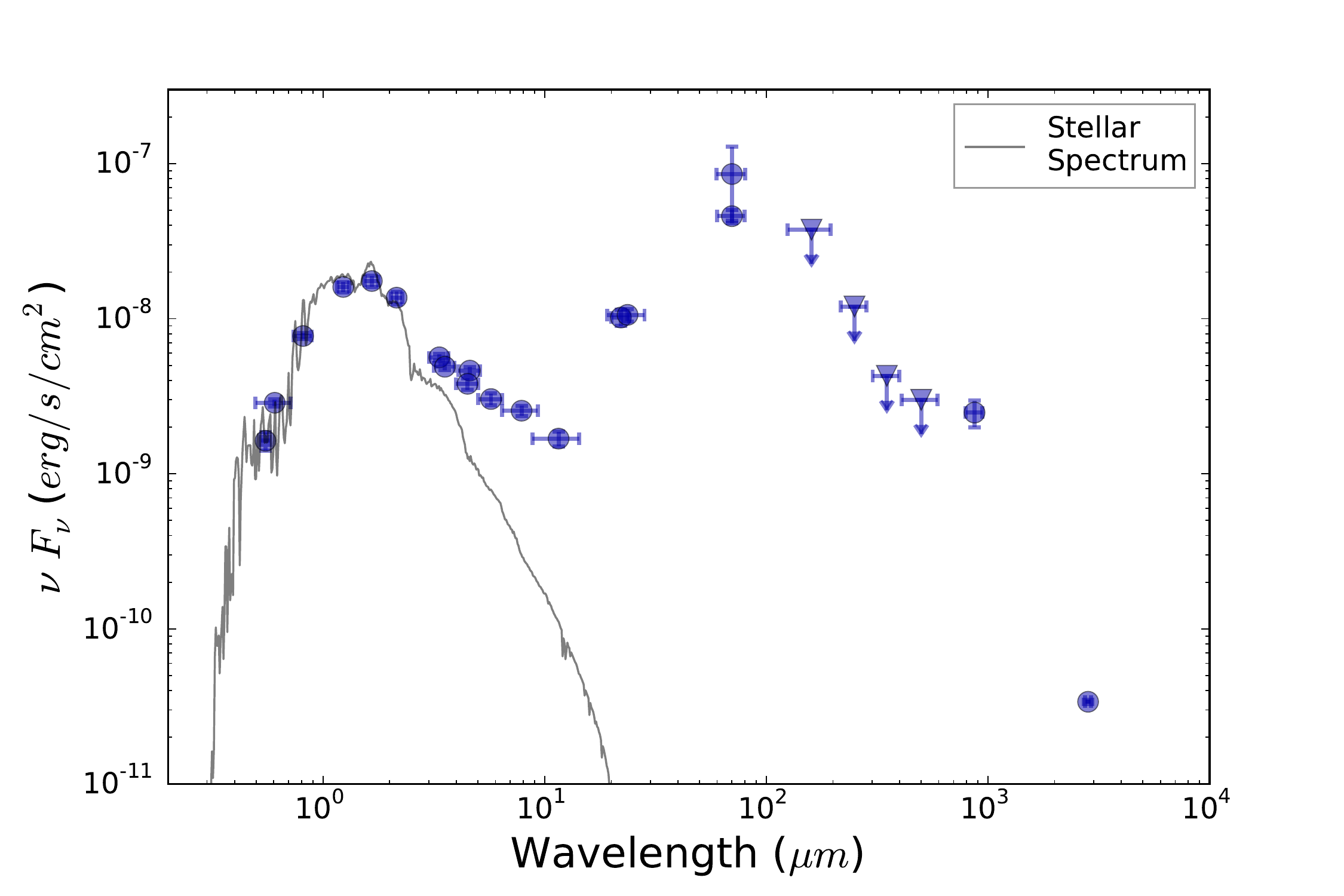}
\caption{Spectral energy distribution for ESO-H$\alpha$ 569 with upper limits indicated by triangles. The SED exhibits the double peaked structure typical of an optically thick, edge-on disk. The values were compiled from the literature with more information given in Table 1. The stellar spectrum for an M2.5 star with $T_{eff} = 3500$ K is overplotted. 
%\todomp{ for the overplotted stellar spectrum, is any extinction $A_V$  applied?} 
\label{sed}}
\end{center}
\end{figure}

A spectral energy distribution (SED) for the disk was compiled from the literature, including data from HST, 2MASS, Spitzer, WISE, Herschel, ALMA, and the LABOCA instrument on the APEX telescope (see Figure \ref{sed}). Table \ref{table:sed} provides the SED values with photometric errors and references for each value. 
The SED shows the characteristic double-peaked shape of edge-on disks with contributions from both the scattered light from the central star peaking at about 1.5 $\mu$m and the thermal emission from the surrounding optically thick disk  peaking at roughly 70 $\mu$m.
Data at similar wavelengths from different epochs show variability at the $10-20 \%$ level, consistent with variability seen in other young disks \citep{2011ApJ...728...49E, 2012ApJ...748...71F, 2009ApJ...704L..15M}.

ESO~H$\alpha$~569 was imaged with Herschel as part of the Gould Belt survey in the PACS 70 and 160 \micron\ bands and the SPIRE in the 250, 350 and 500 \micron\  bands \citep{2012A&A...545A.145W}. The source is barely detected in the PACS bands, hence the large uncertainties reported by \citet{2012A&A...545A.145W}. There seems to be a very marginally-detected point source in the SPIRE bands ($100\pm100$ mJy at 250 \micron\ and $50\pm50$ mJy at 350 and 500 \micron), but given the coarse angular resolution, it's hard to exclude contamination from dust emission from the surrounding cloud itself. 
Given the low significance of these detections, the Herschel fluxes are not included in our SED fits. 

An ALMA Band 3 continuum (2.8 mm, 106 GHz) measurement was obtained by \citet{2016ApJ...823..160D} with ALMA in a compact configuration that achieved a $\sim$2" beam, and does not resolve the disk. Because the disk is not resolved, the ALMA continuum flux could be contaminated with flux from a remnant envelope. However, there cannot be too much non-disk material present, or it would be too opaque to see the central disk in scattered light in the visible as we do. 
This measurement was published after our initial rounds of disk SED fitting as described below (the grid fit described in section 3.2-3.3, and the $\chi^2$-based MCMC fit in section 3.4.1), but this datapoint has been included in our final SED model fitting (the covariance-based MCMC fit described in section 3.4.2).

In addition to the continuum measurement at 0.55 \micron\ included in Table 1, \citet{2012AJ....144...83R} provide fluxes for ESO~H$\alpha$~569 in HST WFPC2's F631N, F656N and F673N narrow band filters associated with [OI], H$\alpha$, and [SII] emission respectively. The disk is not resolved and the measured fluxes are near the detection limits:  0.21 $\pm$ 0.15 (F631N), 2.4 $\pm$ 0.7  (F656N), and 0.35 $\pm$ 0.12 (F673N) $\times 10^{-16}$  erg  s$^{-1}$ cm$^{-2}$ \AA$^{-1}$. These emission lines are all consistent with the spectrum of \citet{2007ApJS..173..104L}, which shows strong H$\alpha$ emission and [SII] emission. Given the large uncertainties, and the fact the the model is not set up to simulate line emission, these emission lines were not included in the SED fits.

%\begin{center}
\begin{deluxetable*}{l|cc}[htbp!]  % <--- column justification (center/left/right)
\tablecolumns{3}
\tablecaption{Modeled Disk Parameters}
\tablehead{   % column headings
  \colhead{Parameter} &
  \colhead{Grid Values} &
  \colhead{MCMC Values} 
}
\tablewidth{5.0in}
\startdata
Distance (pc) & 160 (Fixed) & 160 (Fixed) \\
Outer Radius (au) & 125 (Fixed) & 125 (Fixed) \\
Min Particle Size ($\mu m$) & 0.03 (Fixed) & 0.03 (Fixed) \\
\hline 
Inclination (degrees) & 60 to 90 & 65 - 90\\  % \\ signals new line
Scale Height (H in au at R=100 au)  & 10, 15, 20, 25 & 5 - 25 \\
Dust Mass (M in $M_{\odot}$)   &  $10^{-4}$, $3 \times 10^{-4}$, $10^{-3}$ & $10^{-5}$ - $10^{-3}$  \\
Surface Density ($\alpha$) & -2.0, -1.5, -1.0, -0.5, 0.0 & -2.0 - 0 \\
Flaring Exponent ($\beta$) & 1.1, 1.2, 1.3, 1.4, 1.5 & 1.0 - 1.5  \\
Max Grain size\tablenotemark{a}
 ($\mu m$) &  100, 1000, 3000 & 100 - 3000 \\
Weight\tablenotemark{b} & --- & 0.3 - 0.7 \\
Grain Porosity & 0.0, 0.25, 0.5 & --- \\
Structure & Disk, Tapered Edge Disk & --- \\
\enddata
\label{table:params}
\tablenotetext{a}{Grain size was kept at a constant value of 100 $\mu m$ for the covariance based MCMC run.}
\tablenotetext{b}{During the $\chi^{2}$ based MCMC run a weighting term was used to describe the relative contribution of the image and SED fits to the log likelihood value of each model.}
\end{deluxetable*}
%\end{center}

\section{Model Fitting}
\label{Sec:model}

The scattered light images and full SEDs together provide a comprehensive dataset for ESO~H$\alpha$~569 against which properties of the central star and surrounding disk can be tested. The disk geometry can be directly measured from the images and the distribution of the dust grains within the disk is traced by the SED and disk morphology. To characterize this system, disk models were constructed to explore parameter space with direct comparisons to the observations. The next section presents the context and challenges for radiative transfer modeling of complex disk structures.

\subsection{Radiative Transfer Modeling and Model Fitting of Circumstellar Disks }

Circumstellar disks are complex objects: mixtures of gas and dust, containing solid bodies from the smallest planestimals to giant Jovian planets, shaped by many dynamical forces across evolutionary states from the youngest protoplanetary disks through transitional regimes to second-generation debris disks. This complexity can now be probed by powerful observational capabilities across the entire electromagnetic spectrum, with especially detailed views provided in the visible by the Hubble Space Telescope, in the infrared by 8-10 m telescopes with adaptive optics and soon by JWST, and in the millimeter and submillimeter by ALMA and other interferometers. In some cases a particular physical property of interest can be directly measured from a given observation, but more typically forward modeling of the data must be performed to derive
constraints on the underlying physics. This is particularly necessary for observations of disks at wavelengths where they are optically thick, which is the case for observations of protoplanetary disks at visual and near-IR wavelengths. 

The general outline of such inference is well known: Start from a model of the system's properties and physics with some number of free parameters. Construct synthetic observables using that model, for instance through Monte Carlo radiative transfer calculations. Then compare the synthetic observables to data in order to constrain the free parameters and draw conclusions about their most likely values and the ranges of uncertainty. This process sounds simple enough in theory, but is often a practical challenge due to several confounding factors, among them the complexity of the underlying physics (which inevitably requires simplifications in the models), the nonetheless high dimensionality of the model parameter space, and the need to confront heterogenous and multi-wavelength observations in order to resolve model degeneracies. 

The current work makes use of the MCFOST radiative transfer code \citep{2006A&A...459..797P}, one of a broad class of class of Monte Carlo Radiative Transfer (MCRT) programs designed to study circumstellar disks %\citep[e.g.][]{2009A&A...497..155M, 2011A&A...536A..79R, 2013ApJS..207...30W}.
\citep[for a review of radiative transfer codes see][]{2013ARA&A..51...63S}.
In short, such a code begins with a numerical model of the physical properties within the disk, such as the density of dust in each grid cell, and the mineralogical composition and size distribution of dust particles.  
It then computes the temperature and scattering source function everywhere in the disk via a Monte Carlo method: photon packets are propagated stochastically through the model volume following the equations of radiative transfer, and information on their properties is retained along their path. The radiation field, and quantities derived from it (for instance temperature, radiation pressure, etc) are obtained by averaging this Monte Carlo information. Observable quantities (SEDs and images) are then obtained via a ray-tracing method, which calculates the output intensities by formally integrating the source function estimated by the Monte Carlo calculations. This approach naturally allows simulation of disk images which are dominated by scattered starlight, thermal emission from the dust, or a combination thereof.

Comparison of the simulated images and SEDs against observations then allows inference about which ranges of model parameters are compatible. There are a couple different approaches to performing such comparisons. 
One option is to compute a grid of models spanning the parameter space of interest  \citep[e.g.][]{2006ApJS..167..256R, 2010MNRAS.405L..26W, 2008A&A...489..633P}.  Bayesian techniques allow derivation of uncertainty ranges around the best fit grid point \citep[e.g.][]{2012ApJ...756..168C}. However, even with hundreds of thousands of models computed, given the high dimensionality of the parameter spaces, each parameter must often be quite coarsely sampled at just a few discrete values, which can limit the results achieved. The grid technique is also computationally inefficient because it blindly allocates equal effort to both the best- and worst-fitting portions of parameter space. As is well known the Monte Carlo Markov Chain (MCMC) paradigm improves on this; the MCMC algorithm allows efficient exploration of parameter space and yields detailed information on parameter posterior probability distributions and correlations.

However, most disk model-fitting efforts to date have concentrated on fitting either SEDs alone \citep[e.g.][]{2010A&A...523A..42H,2016MNRAS.458.1029R} or images or interferometric visibilities alone \citep[e.g.][]{2015ApJ...811...18M,2015ApJ...798..124R,2017A&A...605A..34P}. 
%\todomp{add more example cites} 
This is broadly the case independent of the choice of grid fitting versus MCMC fitting.  
But fits to SEDs alone are notoriously degenerate \citep{2001ApJ...547.1077C, 2015EPJWC.10200007W}, and spatially resolved image data or interferometric visibilities are required in order to place robust constraints on many properties of interest. 
Only a handful of disk studies have successfully and rigorously fit models to heterogenous observables including SEDs and images or interferometric visibilities, but when this has been achieved it has often yielded particularly powerful constraints and detailed insights into disk structures \citep[e.g.][]{ 2008A&A...489..633P,2012A&A...539A..17L,2010ApJ...712..112D,2014A&A...567A..51C,2015A&A...577A..57M,2016ApJ...832..110C}.

Such works have most often used the grid fitting approach rather than MCMC, perhaps due to the increased technical complexity of integrating the MCMC framework with heterogenous observables. A detail -- but an important one in this context -- is that the MCMC approach necessarily assesses a single goodness-of-fit metric which must combine both SED and image data together, such as a sum of $\chi^2$ values from the SED and image (or more generally from any combination of distinct observables). In the case where the best-fitting $\chi^2$ for one observable is systematically much higher than for the other observable(s), the model fitting will be driven by that first observable, and will likely not deliver an adequate simultaneous fit to the others.  
Models must necessarily simplify, and imperfect models lead to correlated systematic residuals that increase the minimum $\chi^2$. Consider for instance attempting to fit a simple axisymmetric model to an eccentric disk. This problem is generally worse for images than for SEDs, because the one-dimensional nature of SEDs collapses much of the parameter space. In other words, the well-known degeneracies of SEDs can hide disk offsets, eccentricities, spiral arms, and other asymmetries that are immediately apparent in sufficiently high resolution images.  As a result, it becomes difficult to develop a good metric that combines both images and SEDs in a well-balanced manner for the purposes of a simultaneous MCMC fit.

To address this difficulty in fitting disk observations, a new method has been developed that explicitly takes into account the covariant and correlated residuals in the image fitting. \citet{2015ApJ...812..128C} introduced this approach in the context of 1D spectral fitting. That approach has been extended to work on heterogenous disk datasets including 2 dimensional images, and use that to implement an MCMC fitting process that balances both the image and SED data for ESO~H$\alpha$~569. 
%This is described in detail in Section 3 below.

%%%%%%%%%%%%%%%%%%%%%%%%%%%%%%%%%%%%%%%%%%%%%%%%%%%%%
%%%%%%%%%%%%%%%%%%%%%%%%%%%%%%%%%%%%%%%%%%%%%%%%%%%%%
%%%%%%%%%%%%%%%%%%%%%%%%%%%%%%%%%%%%%%%%%%%%%%%%%%%%%

\subsection{Radiative Transfer Modeling with MCFOST}
\label{modelintro}

For this work, the MCFOST radiative transfer code  \citep{2006A&A...459..797P, 2009A&A...498..967P} was used to construct SEDs and 0.8 $\mu m$ scattered light images for each of the models. 
The 0.6 $\mu m$ scattered light images were not modelled because the strong jet signature required masking $\ge$ 50\% of the integrated disk flux. 

The selected model assumes an axisymmetric disk with a surface density, $\Sigma$, described by a power law distribution in radius given by $\Sigma = \Sigma_{0} (R/R_{0})^{\alpha}$ where $\alpha$ is termed the surface density exponent and $R_{0}$ is the reference radius of 100 au. In this ``sharp-edged'' model the disk is abruptly truncated at an outer radius $R_{out}$.  In order to achieve a good fit to the diffuse emission above the disk and the disk mass and inclination simultaneously, a ``tapered-edged'' disk model was tested, in which the density $\Sigma$ falls off exponentially with some critical radius $R_{c}$ of material outside of the disk:\footnote{Note that some authors give this equation using the notation $\gamma  = - \alpha$.} 
\begin{equation}
\Sigma = \Sigma_{c} \bigg(\frac{R}{R_{c}}\bigg)^{\alpha} \exp \bigg[\bigg(- \frac{R}{R_{c}}\bigg)^{2+\alpha}\bigg]
%(Hughes et al. 2008) 
\label{Eq:taper}
\end{equation} 
For this work, $R_{c} = R_{out}$. This exponential taper is predicted by physical models of viscous accretion disks \citep{1998ApJ...495..385H}, but observations were not sensitive enough to detect this outer gradual fall-off until \citet{2008ApJ...678.1119H} used this form to model both gas and dust continuum observations in the millimeter. 
It is expected that the small dust grains seen in scattered light should be well coupled with the gas for young disks, suggesting the use of this surface density distribution is justified here. \citep[See also recent work by][for HD 163296 and T Cha respectively]{2016A&A...588A.112G,2017A&A...605A..34P}.
%This model was first employed by \citet{2008ApJ...678.1119H} to fit the shape of disks imaged at millimeter wavelengths and based off a surface density profile developed by \citet{1998ApJ...495..385H}. 
The scale height is also defined as a power law in radius by $H(R) = H_{0} (R/R_{0})^{\beta}$ where $\beta$ is the flaring exponent describing the curvature of the disk and again $R_{0}$ = 100 au.

Several model parameters were held fixed to minimize the degrees of freedom and to save computation time. Values for these parameters were either measured directly from the HST images or taken from the literature. The disk is within the SFR Chamaeleon I (Cha I), therefore we fix the distance to the disk at 160 pc \citep{1997A&A...327.1194W}.  From the angular size of the disk measured above and the distance, we calculate an outer radius of 125 au.  The inner radius was defined by a conservative estimate of the sublimation radius $R_{\mathrm{sub}} = R_{\mathrm{star}} (T_{\mathrm{star}}/T_{\mathrm{sub}})^{2.1} \sim 0.1 \, AU $ where $T_{\mathrm{sub}} = 1600 \, K$ \citep{2006ApJS..167..256R}.

The free parameters in the model are inclination (with 90\degr as edge-on), scale height, dust mass, maximum dust particle size, dust porosity, disk vertical flaring exponent ($\beta$), surface density exponent ($\alpha$), and disk edge type (sharp or tapered). For the maximum particle size, the grain population is described by a single species of amorphous dust of Olivine composition
\citep{1995A&A...300..503D} with a particle size distribution following a -3.5 power law extending from 0.03 $\mu$m up to the free parameter $a_{max}$. 
We assume that the dust is well mixed with the gas, irrespective of the particle size.
This combination of dust properties (with $a_{max} = 100 \, \mu m$) results in a mean scattering phase function asymmetry factor of $g = 0.54$. 
Dust porosity is modeled simply as a fraction between 0 and 1 of vacuum that is mixed with the silicates following the Bruggeman effective mixing rule.

For comparison with the observed 0.8 $\mu m$ scattered light images, each model image was convolved with a Tiny Tim simulated PSF \citep{1995ASPC...77..349K}. The 0.8 $\mu m$ observations were masked to select only the pixels with flux values $\geq 3 \sigma$ above the background noise level. 
%Error maps were constructed by adding in quadrature the photon noise ($\sqrt{N}$) and sky background and used to weight each pixel. 
A 2D map of the noise was generated by converting the observed image to electrons, and assigning $\sigma = \sqrt{N_{e-}}$ for the $\chi^{2}$ values.
The model images were aligned with the observations via a cross correlation and normalized to the total observed flux. The models were then compared to the data via an error-weighted pixel-by-pixel $\chi^{2}$ calculation. For similar work see \citet{2010ApJ...712..112D} and \citet{2011ApJ...727...90M}. For the SEDs, when fitting each model point, the foreground extinction is allowed to vary from $A_{V} = 0 \, - \, 10$ with $R_{V} = 3.1$, and the extinction value that minimizes the observed - model residuals is chosen. 

While the robust treatment of radiative transfer provided by MCFOST
%including multiple scattering events
is essential for modeling optically thick disks, it is computationally intensive. Generating a single model SED requires $\sim$ three minutes of desktop CPU time, with an additional $\sim$ minute to generate synthetic images at each desired wavelength. MCFOST allows the user to parallelize the computation, however, systematic explorations of parameter space can quickly become very time consuming.

This complex parameter space was explored in two stages using two different techniques. First, a coarse model grid was computed with a wide range of allowed model parameter values to get a handle on reasonable regions of parameter space. Section \ref{Sec:grids} describes the initial exploration of parameter space via a grid search, with results in Section \ref{Sec:gridres}. This work was used to inform a more robust Markov Chain Monte Carlo exploration for finer sampling of allowed parameter values, with methods described in Section \ref{Sec:mcmcs} and results given in Section \ref{Sec:mcmcres}.

\subsection{Initial Exploration of Parameter space via grid search}
\label{Sec:grids}

Our initial modeling used a uniform grid sampling, with the explored parameter space shown in Table \ref{table:params}. 
For each set of disk model parameters, 15 disk inclinations were sampled uniformly in $\cos{i}$ between 60 and 90 degrees. This resulted in a grid of over 200,000 models. Comparison with data were performed using custom IDL software. A benefit of the grid search approach is that multiple goodness-of-fit metrics may be evaluated across all sampled points. $\chi^{2}$ values were computed separately for the 0.8 $\mu m$ image and SED for each model along with the combined total $\chi^{2}_{tot} = \chi^{2}_{0.8 \, \mu m} + \chi^{2}_{SED}$. 
Bayesian probabilities are derived from the likelihood function wherein the $\chi^{2}$ value for a given model with unique parameter values is related to a probability $\exp(-\chi^{2}/2)$ and the sum of all probabilities is normalized to unity \citep[e.g.][]{2008A&A...489..633P}.

The grid sampling is a simple way to explore parameter space initially, but its sampling of parameters proved to be inadequate for several reasons. First it is too sparse to provide clear insight into degeneracies between the various parameters.  Secondly, the discrete sampling limits the precision with which best-fit values can be determined, and does not allow rigorous computation of uncertainties. These factors motivated the later development of our MCMC model-fitting toolkit described below. Nonetheless the results of the grid search helped clarify relevant portions of parameter space and informed our understanding of the disk.

 \subsection{Results and Conclusions from Grid Search}
 \label{Sec:gridres}

%The best fit parameter values are shown in Table \ref{table:bestfit}. 
For the grid search approach, the best fit model for the disk was found using a tapered-edged disk with non-porous grains, an inclination of 75.5 degrees, and a scale height of 20 au at a reference radius of 100 au. The preferred maximum particle size is 3000 $\mu m$, the dust mass is $3 \, \times 10^{-4}$ $M_{\odot}$, the flaring exponent $\beta$ is 1.3 and the surface density exponent $\alpha$ is -0.5. The separate SED and image fits for the $\alpha$ and $\beta$ exponents favor opposing extremes of parameter space, but the combined $\chi^{2}_{tot}$ likelihood distribution peaks in the middle at physically reasonable values. 

\begin{figure}[hbpt!]
\begin{center}
\includegraphics[trim={0 3cm 0 3cm},clip,width=3.2in]{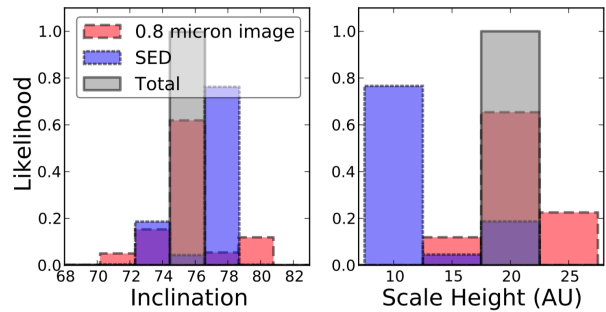}
\caption{Likelihood distributions from the grid search for the disk inclination and scale height computed from the model $\chi^{2}$ values for the 0.8 micron image (red), the SED (blue) and for the combined dataset (grey). The image and SED results favor different regions of parameter space. The sampling of the grid approach is sparse and does not provide an adequate estimate of the uncertainties. 
\label{fig:dists}}
\end{center}
\end{figure}

Figure \ref{fig:dists} illustrates the likelihood distributions for the inclination and scale height. The sparse sampling and disagreement between the model parameters preferred by the image and SED (most pronounced in the scale height) demonstrate the limitations of the grid fitting approach.

\subsubsection{Porosity}

Porous grains were initially included in the modeling parameters to provide a better fit to the flux ratio between the top and bottom disk nebulae. Porous grains are generally more forward scattering, which would increase the flux ratio without needing to increase the line-of-sight inclination.
However, the SED fitting strongly favored non-porous grains. A porosity of $\gtrsim$ 0.5 produced a strong dip in the SED around the 10 - 20  $\mu m$ silicate feature that was not observed for this target.
The overall SED+image fits also favor the non-porous grains, though not as strongly as the SEDs alone. The flux ratio issue was subsequently solved by invoking a tapered edge surface density model for the disk structure. For subsequent modeling, only non-porous grains were used.

\subsubsection{Disk Structure: Sharp vs. Tapered Outer Edge}
\label{Sec:tapervssharp}

When modeling the disk with a sharp outer edge, the SED and image fits preferred very different regions of parameter space. Specifically, it was difficult to simultaneously fit the flux ratio between the top and bottom nebulae of the disk, the diffuse emission above the plane of the disk, and the shape of the disk. Because the disk is not precisely edge-on, the scattering angles differ between the upper and lower disk nebulae. Therefore, changes in the scattering phase function of the grains will change the peak-to-peak flux ratio. 
Any parameter that would increase the flux ratio and emission above the disk (for example increasing the inclination or porosity of the grains) caused too much forward scattering and allowed too much of the light from the central star to appear in the peak. 
Similarly, the diffuse emission above the disk could not be described well by a low mass spherical envelope.

\begin{figure*}[hbpt!]
\begin{center}
\includegraphics[width=6.5in]{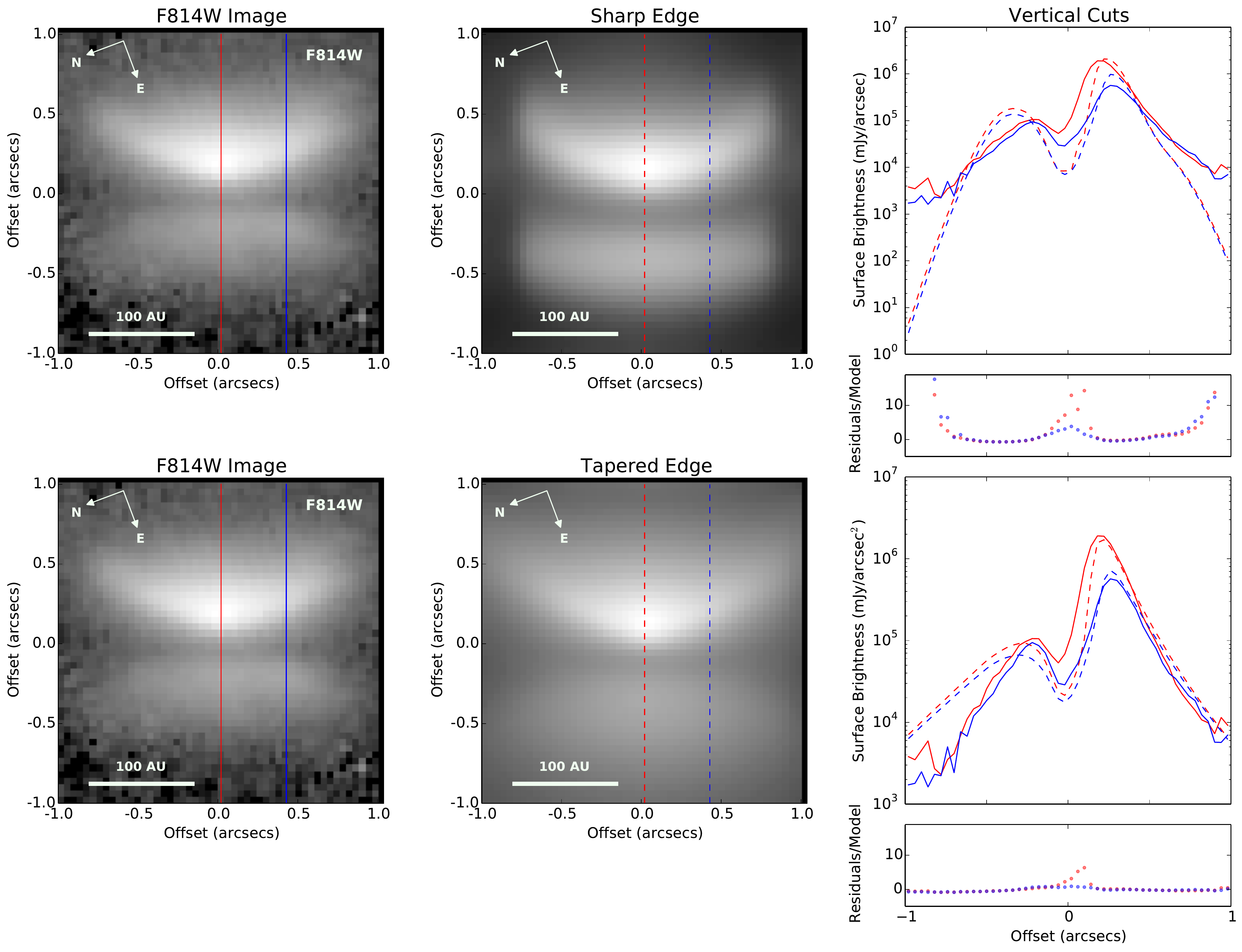}
\caption{Surface brightness profiles for two vertical image cuts through the data (left) and through the sharp-edged (Top middle) and tapered-edged (Bottom middle) disk models. The residuals for the two models are plotted on the same scale (smaller panels at right). The tapered edge model does a much better job of fitting the shape of the disk, especially the depth of the disk midplane and the diffuse outer regions. \label{fig:cuts}}
\end{center}
\end{figure*}

The tapered-edged disk did much better in accounting for both the emission above the disk and matching the flux ratio between the top and bottom sides of the disk. This is clearly demonstrated in Figure \ref{fig:cuts} which compares the observations to the best fit tapered-edged disk model and corresponding sharp-edged model. The right panel shows the surface brightness profiles through several vertical cuts across the disk for both the sharp- and tapered-edged models.

 \subsection{Model Optimization via MCMC}
 \label{Sec:mcmcs}
 
 To more efficiently sample parameter space and gain a better understanding of the uncertainties, a Monte Carlo Markov Chain approach was applied to the model optimization. We used the Python package \texttt{emcee} \citep{2013PASP..125..306F} which implements the Affine Invariant Markov Chain Monte Carlo (MCMC) algorithm by \citet{goodmanandweare}. Specifically, we selected the parallel-tempered MCMC ensemble sampler designed to improve convergence in degenerate parameter spaces. The MCMC samples the posterior distribution given by: 
 \begin{equation}
   \mathrm{P}(\Theta \vert \mathrm{D}) \propto \mathrm{P}(\mathrm{D} \vert \Theta) \mathrm{P}(\Theta) 
 \end{equation}
 %P($\Theta$ $\vert$ D) $\propto$ P(D $\vert$ $\Theta$) P($\Theta$) 
 where D represents the observations, and $\Theta$ the free parameters in the model. Here P(D$\vert \Theta$) is the likelihood of the data given the model and P($\Theta$) is the prior distribution. Uniform priors were adopted for each parameter over the allowable range.

In order to implement this code in conjunction with the MCFOST radiative transfer code, we developed a suite of software tools in Python to interact with the observations, generate models and calculate  goodness-of-fit metrics to inform the MCMC iterations. 
The toolkit is general enough to be usable with any disk image, provided a PSF and uncertainty map are available.
By combining the detailed modeling capabilities of MCFOST with the efficient parameter space sampling of the \texttt{emcee} package, the goal was to self-consistently and simultaneously fit a wide variety of observables in order to place constraints on the physical properties of a given disk, while also rigorously assessing the uncertainties in the derived properties. The \texttt{mcfost-python} package is publically available on github\footnote{https://github.com/swolff9/mcfost-python}, and the authors encourage its use by the disk modeling community \citep{mcfost-python}.

The \texttt{mcfost-python} package was designed to be modular, with different components to read in the observables, interact with the MCFOST parameter files, generate model SEDs and images, compare them to data, and setup and control the overall MCMC run. To validate the functions for comparing models to data, benchmark cross-checks were performed to compare the new Python fitting code to existing $\chi^{2}$ routines in IDL and Yorick. While this code was originally designed to work with HST data and the MCFOST modeling package as described in this paper, it has also been expanded to work with data from different instruments, including polarimetry data, and can be used with other radiative transfer modeling codes.

\subsubsection{$\chi^{2}$ Based Log-Likelihood Estimation}

The \texttt{mcfost-python} package allows the user to choose between two goodness-of-fit metrics. This section discusses the first of those, the $\chi^{2}$ metric. 
A simple benchmark comparison of the $\chi^{2}$ and covariance likelihood methods is provided in the Appendix.
At each step in the MCMC iteration, a model image and SED are created for the chosen parameter values and a $\chi^{2}$ value is calculated using the same methodology as the grid sampling approach. The \texttt{emcee} code requires a log likelihood distribution which is computed from the $\chi^{2}$ assuming a multi-dimensional Gaussian likelihood function:
\begin{equation}
\ln[P(D| \Theta)] = -0.5 N \ln{2 \pi} + \sum_{i=1}^{N} (- \ln{|\sigma_{i}|}) - \frac{1}{2} \chi_{i}^{2}
\label{eq:likelihood}
\end{equation}
Here N is the number of data points, and $\sigma$ is our uncertainty.
The MCMC approach inherently requires a single goodness-of-fit metric, so it is essential to combine the SED and image metrics into a single log likelihood function for use by \texttt{emcee}. The log likelihood distribution is computed separately for the images and SEDs, and a weighted average is used to determine the goodness of fit.

During initial tests using the $\chi^{2}$-based log-likelihood goodness of fit metric, we chose to allow the relative weighting between the image and SED to vary. The best way to handle relative weighting between different types of observations for a single disk model was not well understood, and is a nuisance parameter that does not, itself, inform us about any inherent physical properties of the disk. By marginalizing over it in this way, the intent was to produce a best fit model that was informed by both the SED and image data without a bias towards one or the other. The weighting was allowed to vary between 0.3 and 0.7 for a minimum of 30\% weighting to either the image or SED fits. We found that the image likelihood values were down-weighted due to their systematically higher $\chi^{2}$ values, and the MCMC chains worked to improve the images while largely ignoring the better SED fits. 
In our first round of MCMC calculations, the image reduced $\chi^{2}$ values tended to be more than an order of magnitude above the SED reduced $\chi^{2}$ values (best $\chi_{SED}^{2} = 1.3$, $\chi_{0.8\mu m}^{2} = 66$), due to the larger number of measurements in the images presumably with under-estimated uncertainties.

\subsubsection{Covariance Based Log-Likelihood Estimation}

The imbalance between the image and SED $\chi^{2}$ values served as the impetus for the development of the covariance matrix likelihood estimation method, which ultimately provided much better relative weighting of the different observables.
Given that each model image is convolved with an instrumental PSF, neighboring pixels must be covariant. Furthermore, this approach lets us correct for the global limitations of the disk model to fit the dataset. Model systematics present as correlated uncertainties. For a more complete estimate of the errors in our HST images, we adopt and extend the covariance-based method for log likelihood estimation presented by \citet{2015ApJ...812..128C} in the context of 1D spectral fitting. That approach must be extended to work in the context of 2D images. In this case, we convert Eq. \ref{eq:likelihood}, which describes the likelihood of the data given the model assuming a Gaussian likelihood distribution, into a matrix formalism in Equation \ref{eq:likelihood_matrix}.
\begin{multline}
\ln[\mathrm{P}(\mathrm{D}| \Theta)] = -\frac{1}{2} (R^{T} C^{-1} R + \ln[\det(C)] \\ + N \ln[2 \pi])
\label{eq:likelihood_matrix}
\end{multline}
where $R$ represents the residuals of the observations subtracted by the model, $C$ is the covariance matrix defined below, and $N$ is the total number of pixels in the image (\textit{not} the number of pixels along a given dimension of the array).  

To apply this approach, each 2D image must first be "unwrapped" into a 1D array. In practice not all pixels in an image may have a sufficient SNR disk detection to justify fitting. Excluding such pixels from the unwrapping improves the overall computational efficiency, particularly for the matrix inversion calculation, at the cost of somewhat more complex bookkeeping between the 2D and 1D versions of the image. 

The covariance matrix $C$ (of size $N_{pix} \times N_{pix}$) incorporates both the noise in each individual pixel and global covariances between adjacent pixels (represented by $K^{G}$): $C_{i,j} = \delta_{i,j} \sigma^{2}_{i,j} + K^{G}_{i,j}$. An example source of global covariance is the FWHM of a telescope PSF. For a non-zero PSF FWHM, neighboring pixels cannot be treated as individual measurements of the disk surface brightness. Additionally, any global limitations of the model to fit the data can be implicitly included in the covariance structure. 
For example, when using a symmetric disk model any asymmetries in the observed image of the disk will necessarily lead to higher correllated residuals even for the best-fitting model parameters. These residuals will in general be spatially correlated on one or more scales from the angular resolution to the size of the observed asymmetry. Incorporating our knowledge of these residuals in the covariance matrix improves our ability to draw conclusions given such necessarily imperfect models. Likewise, the choice of incomplete or simplified  parameterizations of the disk physics/structure in the model can be handled the same way. For instance, if there exists an additional un-modeled component such as a more vertically-extended disk atmosphere or significant residual jet emission on the top/bottom on the disk, or if the functional form of the power law adopted for the disk surface density is an oversimplified description of the true disk properties, such systematics would lead to correlated residuals in data-model comparisons. This covariance framework allows the down-weighting of these contributions within the correlated residuals without masking them altogether.

The field of Gaussian processes has developed several useful analytic models for convolution kernels that can be used to parameterize covariant structure. For instance \citet{2015ApJ...812..128C} adopt the Mat\'ern kernel truncated by a Hann window function. This kernel has several free parameters, which can be solved for as nuisance parameters as part of the MCMC fit. Of course, this increases the dimensionality of the parameter space that must be explored, which can in practice increase computation time by an order of magnitude or more. \citet{2015ApJ...812..128C} note that, because the best fit model parameters are relatively insensitive to the precise values of the covariance parameters (i.e. a reasonably good but perhaps not optimal covariance model often suffices), one can first roughly optimize the covariance model and then perform the MCMC fit with that model fixed.  Given the computational demands of disk radiative transfer model fitting, a variant of that approach is adopted here.

\begin{figure}[hbpt!]
\begin{center}
\includegraphics[width=3.2in]{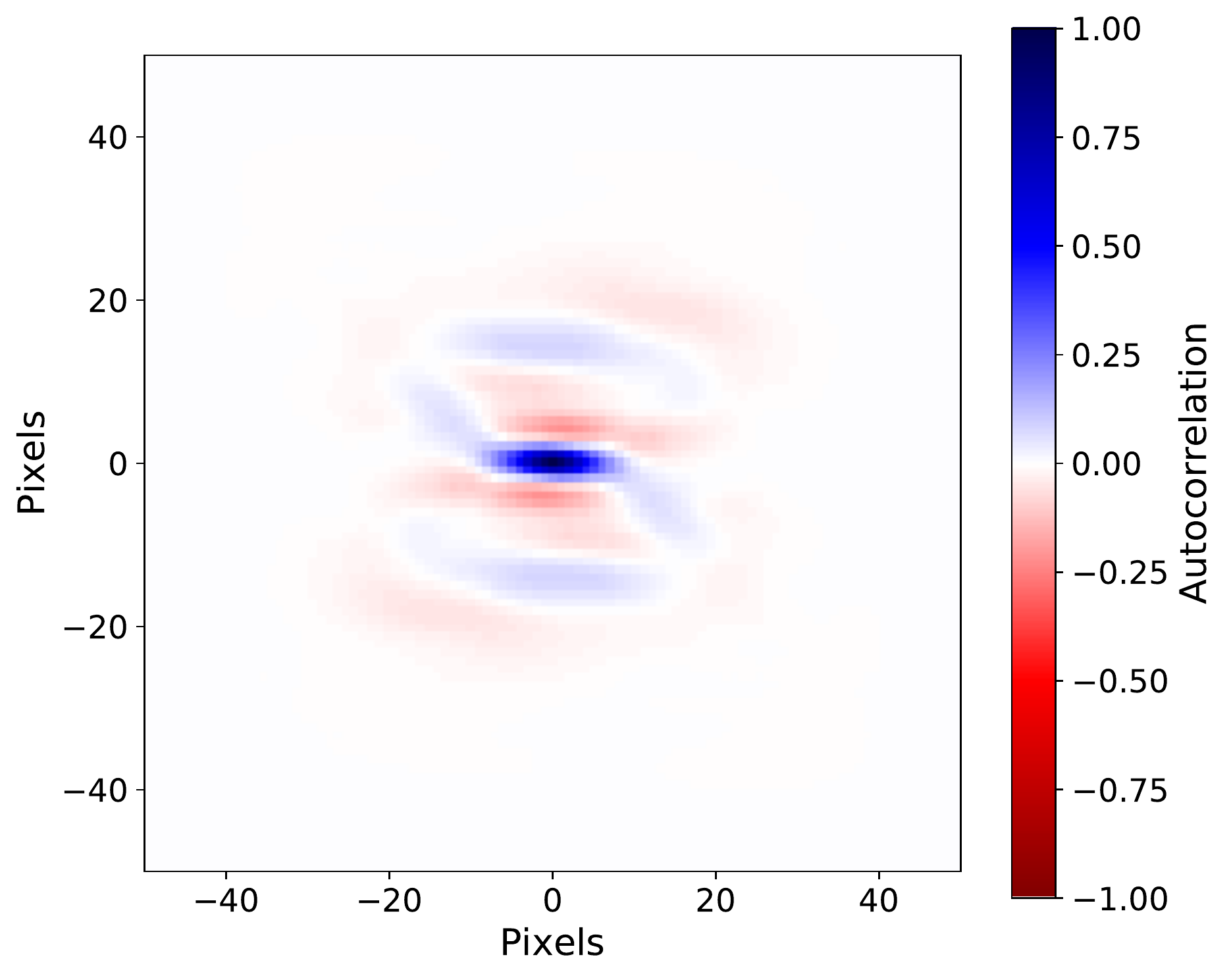}
\caption{The mean of the autocorrelation of the residuals from subtractions between our 0.8 $\mu m$ observed image and a randomly selected subset of 1000 model images spanning the range of the priors. Residuals are most strongly correlated between pixels that are horizontally adjacent, as expected for an edge-on disk with its major axis oriented horizontally. The slight anti-correlation in the vertical direction is likely due to dark lane structure between the two bright lobes. \label{fig:autocorr}}
\end{center}
\end{figure}

\begin{figure}[hbpt!]
\begin{center}
\includegraphics[width=3.2in]{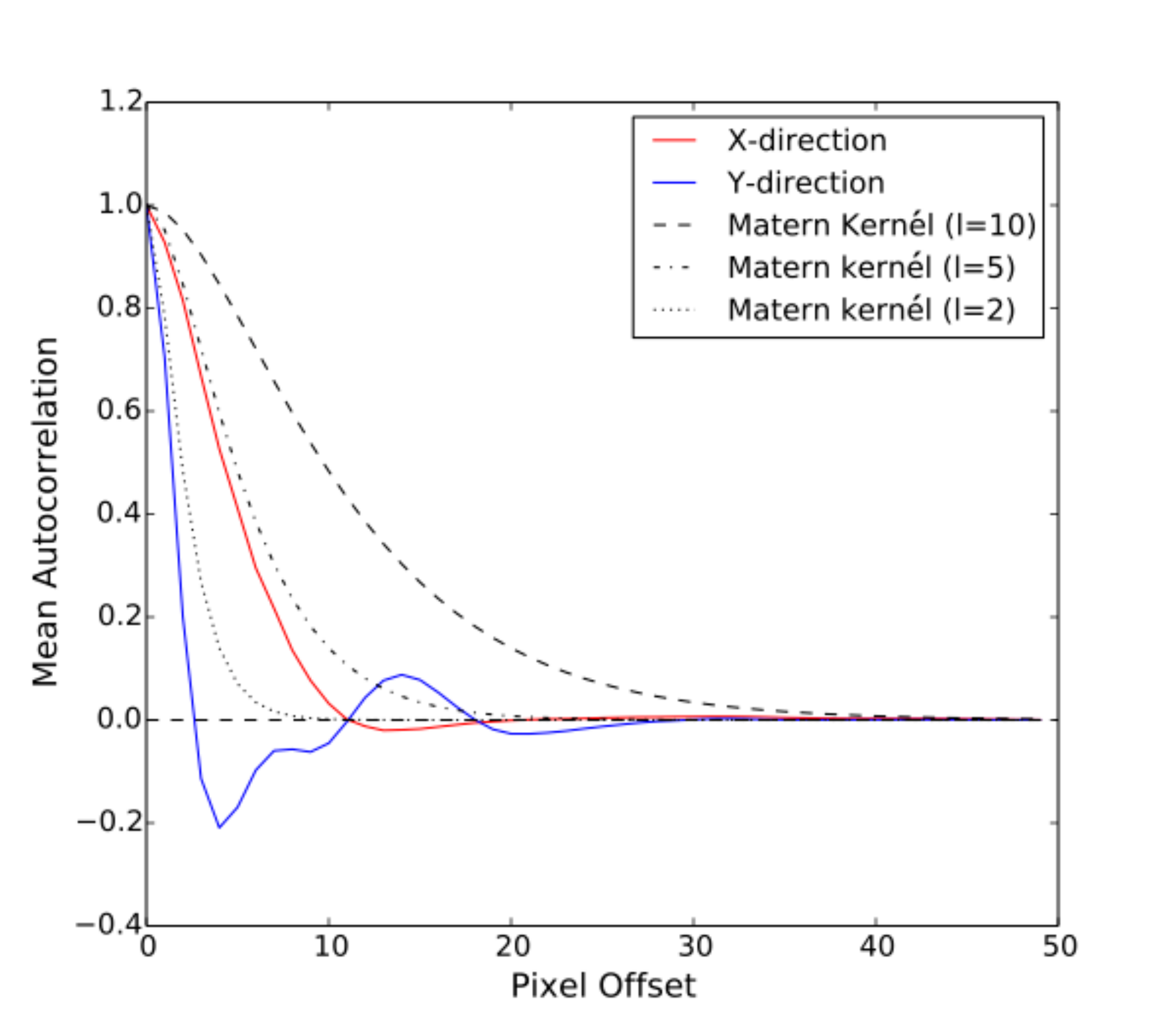}
\caption{Slices through the mean autocorrelation shown in Figure \ref{fig:autocorr}. Both the vertical (blue) and horizontal (red) slices along with several Mat\'{e}rn kernels are presented for comparison.
%The horizontal slice is used to generate our global covariance matrix. 
We conservatively adopt the wider correlation scale from the horizontal axis to generate the global covariance matrix. 
It is not unsurprising that the autocorrelation image is more broadly extended in the horizontal direction where the disk is elongated than in the vertical where the gradients in the disk are much sharper.  \label{fig:1dautocorr}}
\end{center}
\end{figure}

\begin{figure}[hbpt!]
\begin{center}
\includegraphics[width=3.2in]{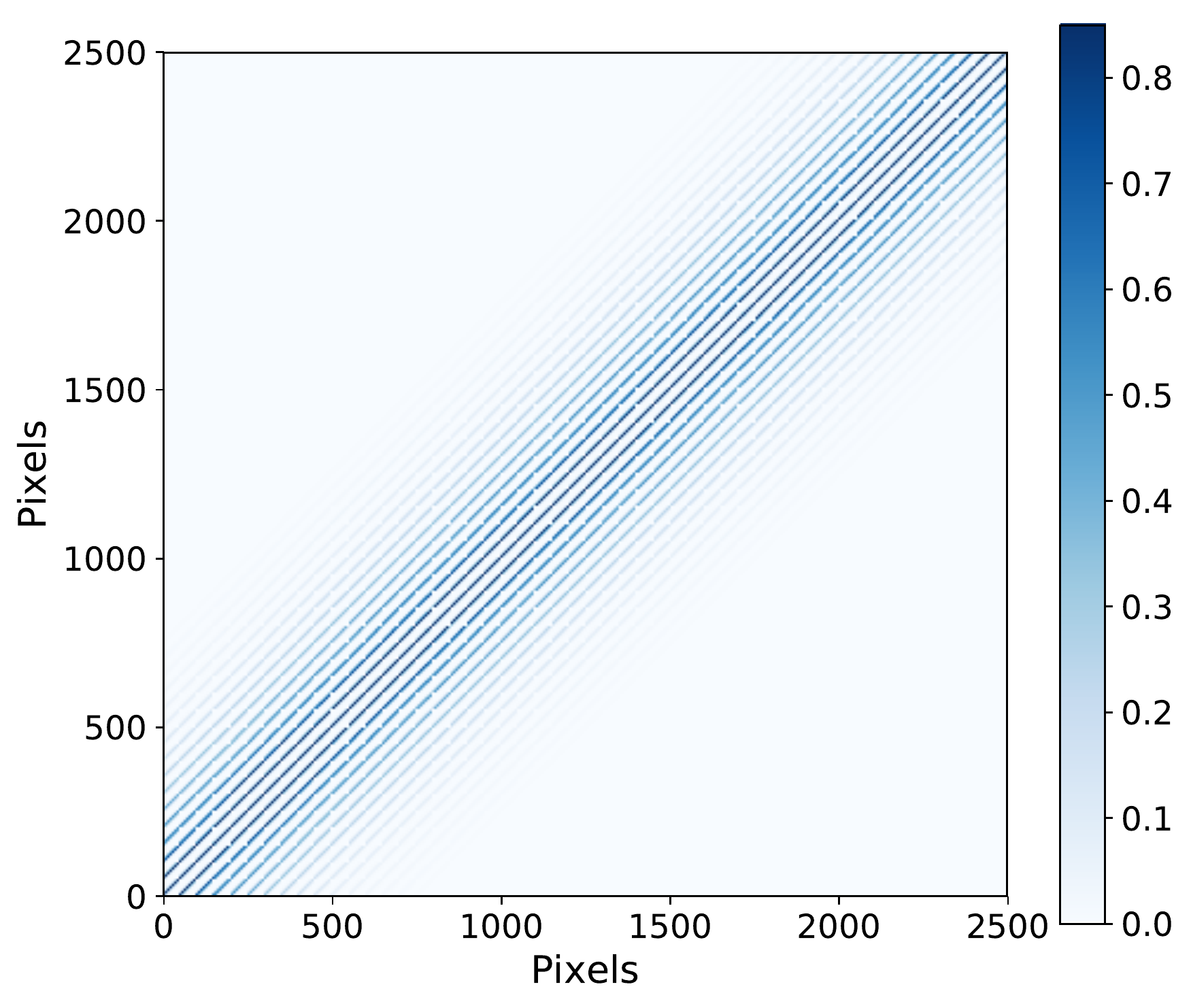}
\includegraphics[width=3.2in]{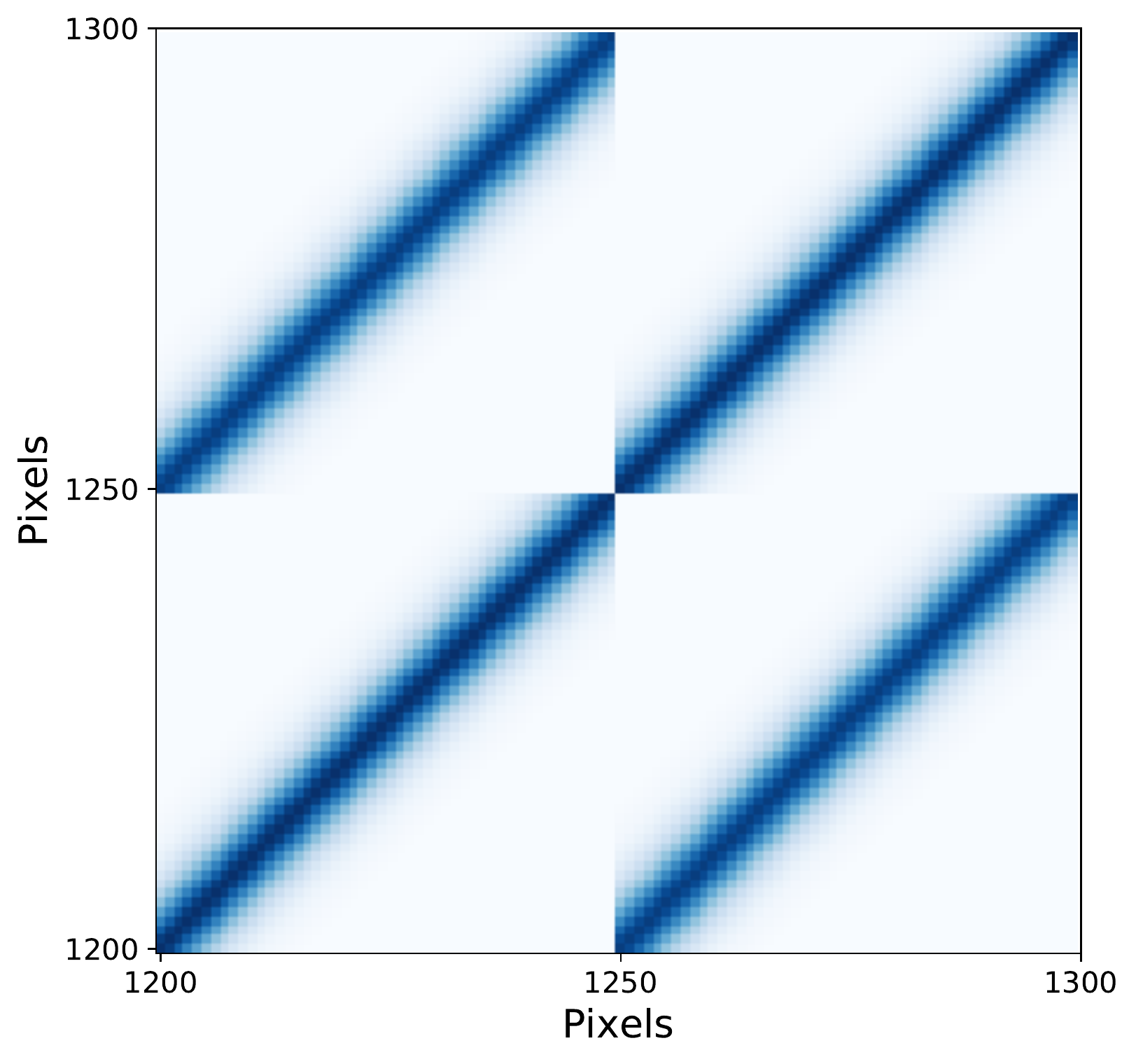}
\caption{Top: Covariance Matrix ($C_{i,j}$) used to compute the log likelihood of the model images given the observations. The matrix combines information about the noise in the observations, the covariances between adjacent pixels, and the pixel mask. Botom: a zoomed in region illustrating the contribution of the autocorrelation function between adjacent pixels. 
To generate this, the 2D 50 $\times$ 50 pixel image is first unwrapped into a 1D 2500 pixel array by stacking each row horizontally. The diagonal of the covariance matrix gives the uncertainties associated with each pixel (where $i=j$). The other elements of the covariance matrix dictate the covariances between the various pixel pairs ($i,j$) which is given by the autocorrelation shown in Figure \ref{fig:1dautocorr} and depends on the distance between pixel $i$ and $j$ in the 2D detector frame (not in the 1D unwrapped image). } \label{fig:covariance}
\end{center}
\end{figure}

The global covariance is estimated empirically by computing the average autocorrelation of the residuals from a subtraction of our 0.8 $\mu m$ image and a subset of 1000 randomly chosen model disk images from a uniform sampling of the parameter space within the limits of our priors (Figure \ref{fig:autocorr}). This provides, in a computationally tractable way, a reasonable model for the covariant structure found in residuals for the parameter space of interest, and allows us to hold the covariance model fixed in subsequent MCMC runs. The 2D autocorrelation is collapsed along the horizontal axis to generate a 1D autocorrelation function (Figure \ref{fig:1dautocorr}). The horizontal axis was chosen to generate the covariance matrix because it provided the most conservative estimate, with a wider tail similar to the Mat\'{e}rn kernel, and did not exhibit the anti-correlation found in the vertical axis due to the dark lane of the disk. For comparison, several $\nu = 3/2$ Mat\'ern kernels are shown, following the chosen formalism from \citet{2015ApJ...812..128C}. 
To compute the covariance matrix $K^G$, for each pair of pixels $i,j$, the distance between them given by $r_{i,j} = \sqrt{(x_{i}-x_{j})^{2}+(y_{i}-y_{j})^{2}}$ is calculated. For each entry of $K_{i,j}^{G}$, the analytic autocorrelation function is interpolated to the value for $r_{i,j}$, with a cutoff outside of 20 pixels to make computations of $C_{i,j}$ manageable. The resulting covariance matrix is shown in Figure \ref{fig:covariance}.

For consistency, the likelihood of each model SED is computed using the covariance matrix framework from Equation \ref{eq:likelihood_matrix}. In this case, the covariance matrix contains only the individual uncertainties for each point multiplied by an identity matrix. Any global limitations of the model SEDs to fit the dataset are neglected. Given the low $\chi^{2}$ values achieved for the SED fitting in the grid search described above (lowest SED $\chi^{2}$ $\sim$ 1.3), the uncertainties in the SED are presumed to be well estimated. For this dataset, covariances are not anticipated between neighboring photometric points, but this formalism would naturally handle any such correlations and makes it straightforward to include continuous spectra as part of a unified fit alongside broadband photometry.

The covariance framework is also capable of including model terms for additional regions of locally covariant structure ($K^{L}$), as discussed in \citet{2015ApJ...812..128C}. We leave the application of such local covariances to disk image fitting for future work, along with the exploration of how best to explicitly model covariances between the SED and image portions of the overall fit.

\subsubsection{Choice of parameter values for MCMC}
\label{Sec:MCchoices}

The allowed parameter ranges were adjusted slightly for the MCMC modeling compared to the grid fit. The computation time for the grid modeling depended both on the number of free parameters, and on the size of the allowed parameter ranges, while the MCMC modeling time depended only on the number of free parameters. Therefore, we were able to widen the prior distributions for the MCMC modeling, being careful to widen allowable ranges for those parameters that were best fit at the edges of the grid distribution, such as the scale height and disk mass. Parameter ranges are shown in Table 2, Column 3. During the IDL grid search modeling phase, we found the image and SED fits both prefer a large maximum grain size. In order to limit the computation time in the MCMC fits, the maximum grain size was fixed to be 3000 $\mu m$, and as noted above, the porosity was set at zero. 

One downside to the MCMC over the grid search approach is that the chain does not work well with discrete parameter distributions. For example, the abrupt distinction between the tapered and sharp edged disk models could not have been tested using MCMC. Given the strong support for the tapered edge disk model as described in Section \ref{Sec:tapervssharp} we selected an exponentially tapered outer edge for the MCMC run.

An MCMC run was conducted using the covariance-based log-likelihood goodness of fit metric with 2 temperatures with 50 walkers. Uniform prior distributions were used for all of the parameters (with the dust mass uniformly distributed in log-space). 
The chain was run for $N_{\mathrm{steps}} = 10,000$, with an initial burn-in stage of $N_{\mathrm{burn}} = 0.2 \, N_{\mathrm{steps}}$. This resulted in a total of 21,000 models requiring $\sim$ 2 weeks of computation time parallelized over only 10 cores. This was a significant improvement over the grid search approach which necessitated generating $\sim$ 200,000 models. As a test of convergence, we compute integrated autocorrelation times ($\tau_{x}$) for each of the parameters and use these to estimate the effective sample size, ESS = $N_{\mathrm{samples}}/(2 \tau_{x})$ (a measure of the effective number of independent samples in the correlated chain). The ESS varied from 761 to 12075 with the surface density exponent being the least well constrained parameter. 
The Monte Carlo standard error for each parameter decreases with increasing effective sample size as $\sigma_{i}/\sqrt{ESS}$ where $\sigma_{i}$ is the standard deviation for the posterior distribution \citep[See discussion in][]{2017arXiv170601629S}. For example, to measure the 0.025 quantile to within $\pm$0.01 with a probability 0.95 requires 936 uncorrelated samples (this corresponds to roughly 10\% errors in the best fit parameter values assuming the tail of the posterior is well described by a normal distribution), which is achieved for all parameters except the surface density distribution where  the 0.025 quantile was only confined to within roughly $\pm$0.0125 \citep{bayesianstatistics}.

 \subsection{Results and Conclusions from MCMC}
\label{Sec:mcmcres}

\begin{figure*}[hbpt!]
\begin{center}
\includegraphics[width=2.0in]{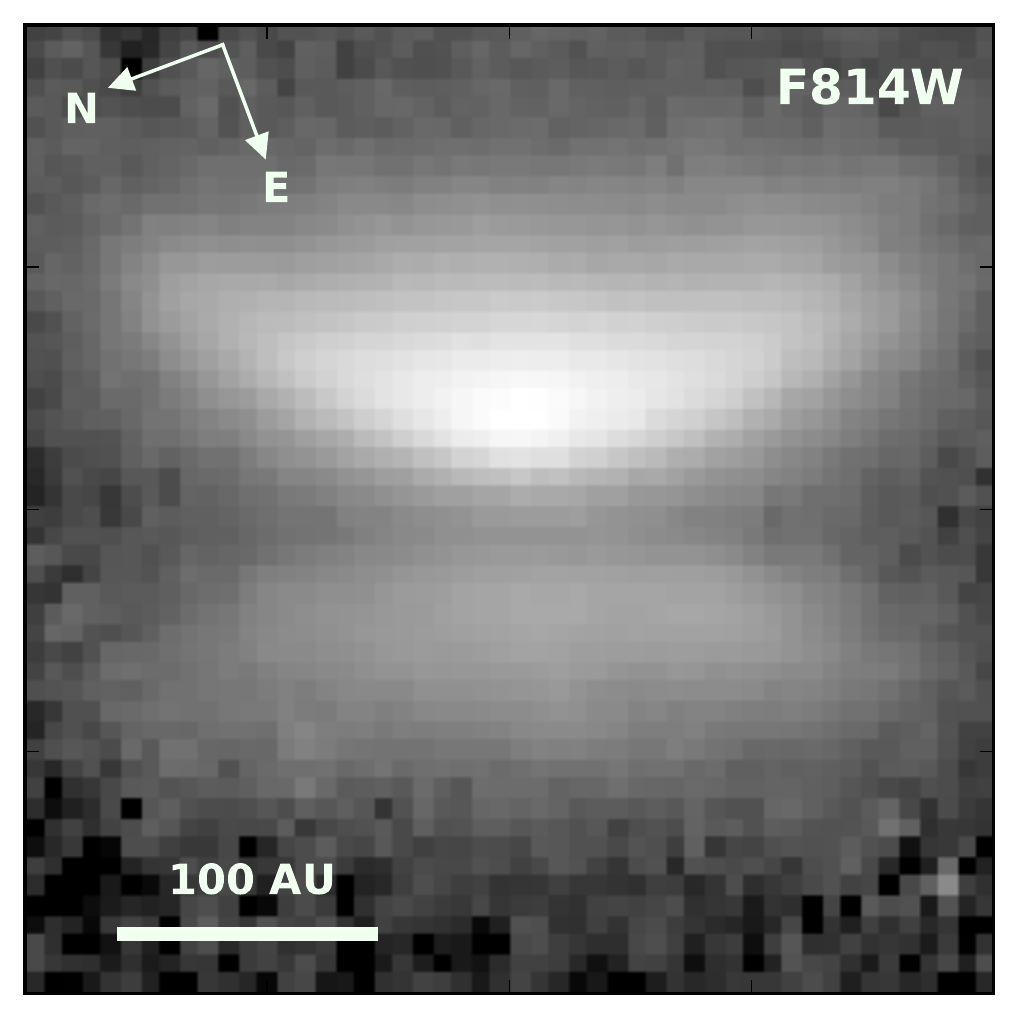}
\includegraphics[width=2.0in]{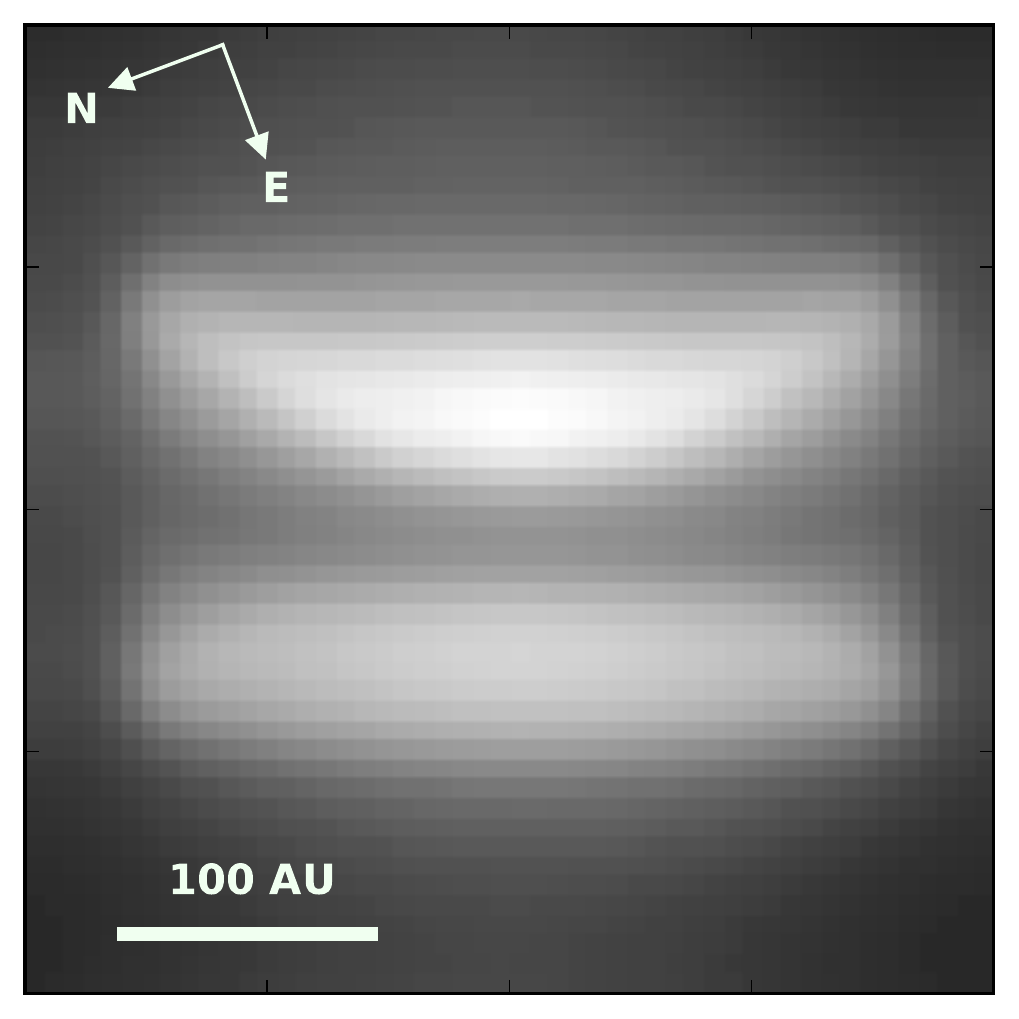}
\includegraphics[width=2.0in]{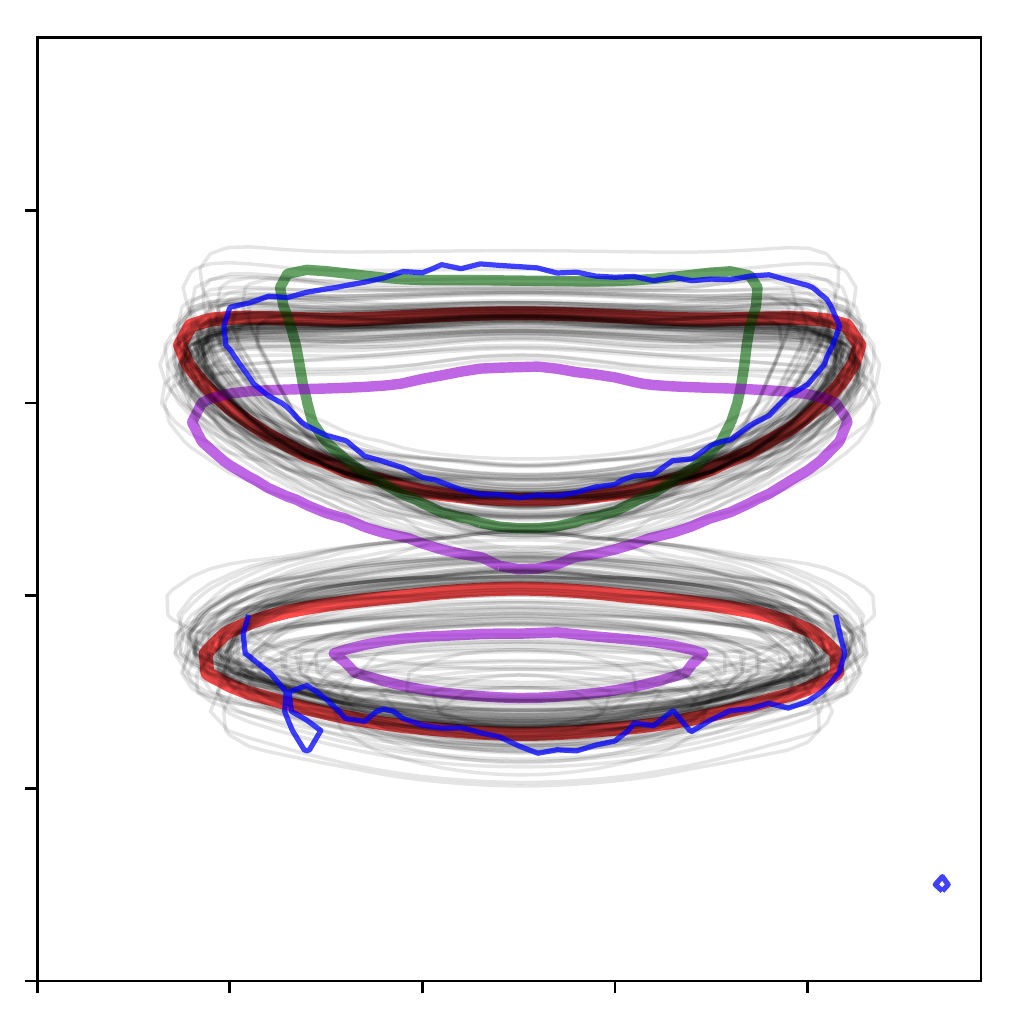}
\includegraphics[width=5.2in]{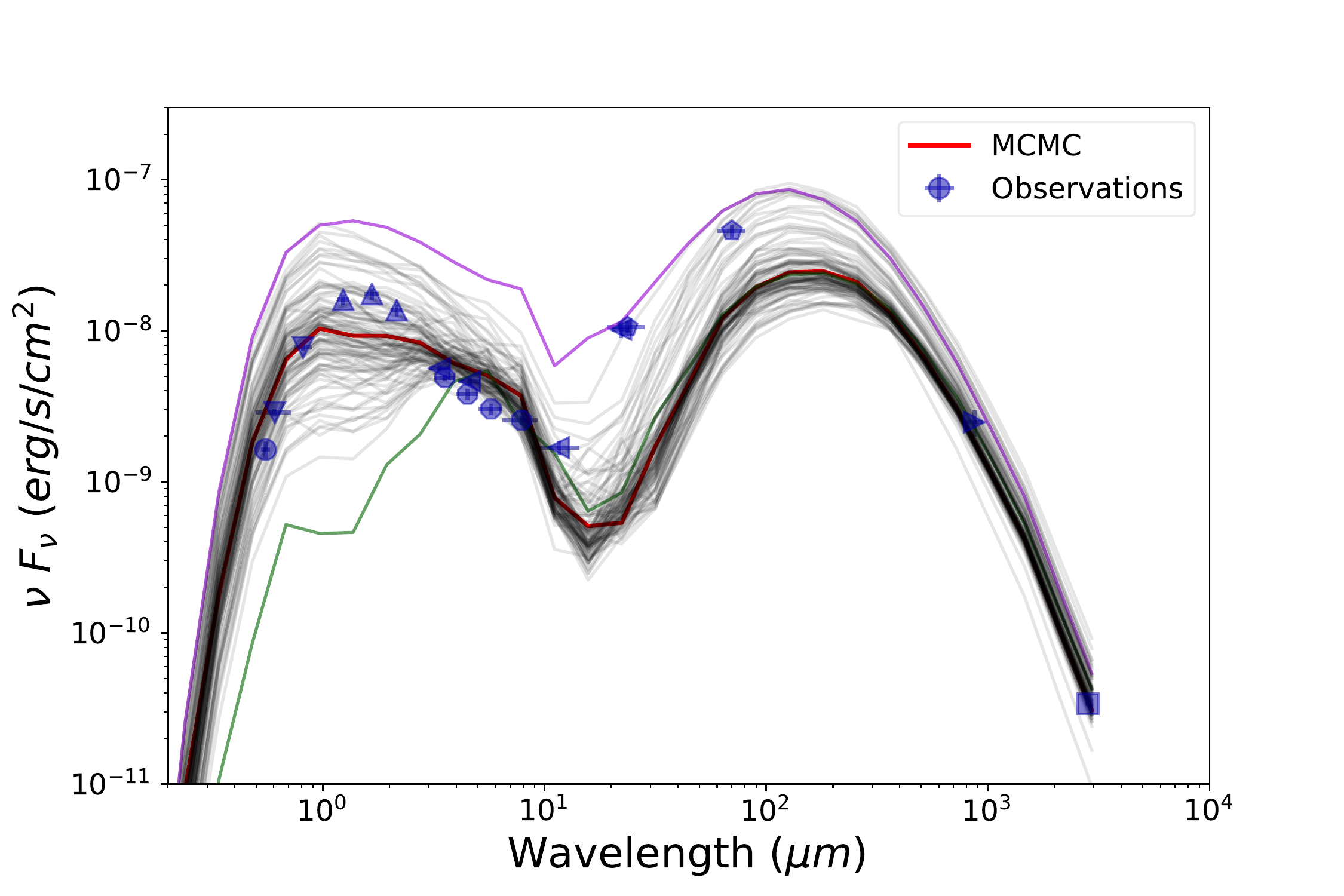}
\caption{The results from the covariance-based MCMC. \textbf{Top:} The model image (Middle) corresponding to the best fit parameters given in Table \ref{table:emceebestfit} compared to the 0.8 $\mu m$ observed image (Left). The right panel shows a contour highlighting the shape of the best fit model disk in red, with contours scaled to the observed 0.8 $\mu m$ image shown in blue. One hundred randomly chosen models drawn from the MCMC chain are depicted in grey. \textbf{Bottom:} The SED for the same model as above is shown in red and compared to the literature values in blue. The grey curves present the same 100 randomly selected models drawn from the chain. While the MCMC results provide a reasonably good fit to both the image and SED, the compromise between the two datasets, inherent in the covariance framework, lead to imperfect solutions. For example the best fit model under-predicts the flux in the 20 - 100 $\mu m$ region. The green and purple lines shown in both the SED and image contours highlight two of the models that are poor fits to the observations. The purple model over-predicts both the flux in the SED and the surface brightness ratio between the top and bottom nebulae in the image. }
\label{fig:bestfit}
\end{center}
\end{figure*}

\begin{figure*}[hbpt!]
\begin{center}
\includegraphics[width=7.0in]{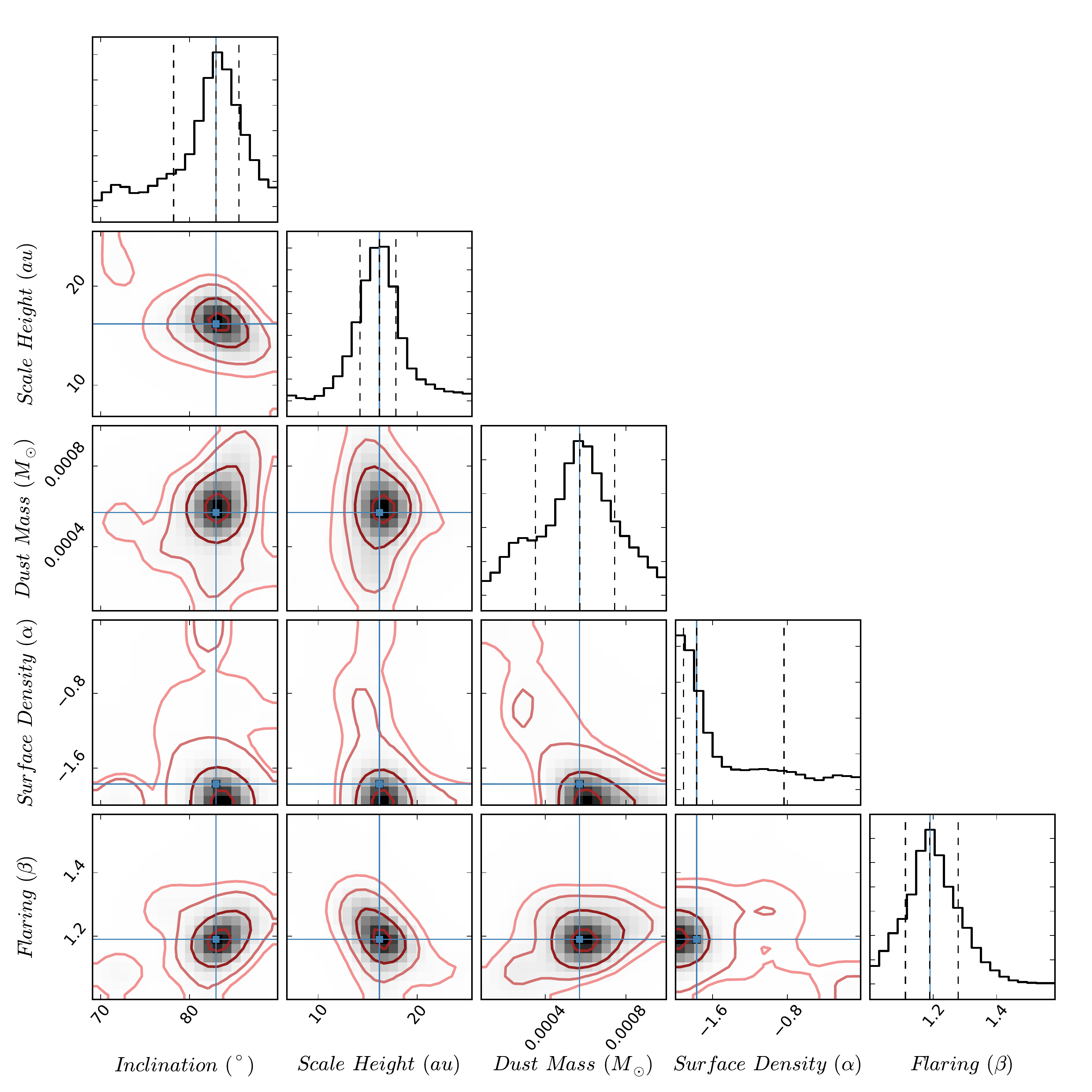}
\caption{The MCMC results using the covariance log likelihood estimation. The blue crosshairs indicate the best fit value for each parameter. Shading indicates the density of the parameter space sampling, while the red contours are drawn at the 1-4$\sigma$ levels. All parameters are well constrained except for the surface density exponent ($\alpha$). Dashed vertical lines represent the 16th, 50th, and 84th percentiles of the samples in the marginalized distributions. \label{fig:distributions}}
\end{center}
\end{figure*}

The best fit parameter values are shown in Table \ref{table:emceebestfit}. 
The data are best fit by a tapered-edged disk with an inclination of $83.0^{+2.6}_{-4.8}$ degrees, a scale height of $16.2^{+1.7}_{-2.0}$ au at a reference radius of 100 au, a disk dust mass of $0.00057^{+0.00017}_{-0.00022}$ $M_{\odot}$ (assuming a gas to dust ratio of 100), a surface density exponent ($\alpha$) of $-1.77_{-0.14}^{+0.94}$, and a flaring exponent ($\beta$) of $1.19_{-0.08}^{+0.09}$. 
%The SED and image distributions for the $\alpha$ and $\beta$ exponents favor opposing edges of parameter space but the total distribution peaks in the middle at physically reasonable values. 
The image and SED combined best fit model is illustrated in Figure \ref{fig:bestfit} (single best fit in red, along with an ensemble of well-fitting models in gray) and together provide a close fit to the observations. Parameter distributions are shown in Figure \ref{fig:distributions}.
%The figure also provides 100 randomly chosen models drawn from the chain. 
The results for individual parameters are discussed in more detail below. 

The best fit parameters provide a compromise between the image and SED fit. Therefore, this combined fit to the SED and image is not as favorable as if the fits had been performed separately on each individual dataset. For example the best-fit model SED under-predicts the flux in the 20 - 100 $\mu$m region of the SED (by a factor of 20 at 20 $\mu$m and 1.5 at 70 $\mu$m), while the best-fit model image under-predicts the flux ratio between the top and bottom nebulae by a factor of $\sim$ 4. Either of these could have been improved individually if we had only optimized the fit for that metric alone.
Models that best fit the image tend to over-predict the disk flux at all wavelengths, while the models that best fit the SED tend to produce images that have very steep surface density profiles, which removes the diffuse material on the outer edges of the disk provided by the tapered edge. 

The apparent disagreement is likely a result of some limitations in the disk model. 
If the opacity of the dust grains in the disk was decreased, the optically thick/thin boundary would move to shorter wavelengths, recovering some of the flux in the several tens of $\mu m$ range of the SED. 
However, to improve the flux ratio between the top/bottom nebulae in the modeled image we would need to move the inclination farther from edge-on and/or change the scattering properties of the grains (i.e. increase the forward scattering or decrease the dust albedo), which would most likely necessitate an increase in the dust opacity.

\begin{deluxetable}{lc}[htbp!]  % <--- column justification (center/left/right)
\tablecolumns{2}
\tablecaption{MCMC Best fit Paramters}
\tablehead{   % column headings
  \colhead{Parameters} &
  \colhead{Best-fit Values} 
 % \colhead{Total}
}
\tablewidth{2.5in}
\startdata
Inclination ($\degr$) & $83.0^{+ 2.6}_{-4.8}$ \\ % & 70.25  &  79.65  \\  % \\ signals new line
Scale Height (au)  & $16.2^{+1.7}_{-2.0}$ \\ %& 23.9 &  13.98  \\
Dust Mass ($M_{\odot}$) &  $0.00057^{+0.00017}_{-0.00022}$  \\ %& 6.861 $ \times \,10^{-4}$ &  8.587 $ \times \,10^{-4}$  \\
Surface Density $\alpha$ & $-1.77^{+0.94}_{-0.14}$\\ % & 1.495  & 1.192   \\
Flaring $\beta$     & $1.19^{+0.09}_{-0.08}$ \\ %& -1.783 &  -1.224  \\
%Max Grain size ($a_{max}$) & $1696^{+786}_{-730}$ % & 749.9  &  376.3  \\
%Weight & $0.34^{+0.12}_{-0.04}$ & 0.3997 & 0.572 
\enddata
\tablecomments{Best fit values for the covariance likelihood estimation mode of the MCMC.}
\label{table:emceebestfit}
\end{deluxetable}

\subsection{Dust Mass}
\label{Sec:mass}

The best-fit disk dust mass is $0.00057^{+ 0.00017}_{- 0.00022} \, M_{\odot}$, which corresponds to a disk mass of 0.057 $M_{\odot}$ (assuming the standard ISM gas to dust ratio of 100). This is 16\% of the stellar mass for a 0.35 $M_{\odot}$ star like ESO~H$\alpha$~569, a surprisingly high disk to star mass ratio. 

In the grid fit and the initial MCMC run using the $\chi^2$ estimator, the fit to the disk mass relied heavily on the 870 \micron\ measurement. 
Given the surprisingly high mass estimate, we speculated that the 870 \micron\ photometry might be in some way compromised (for instance, contaminated by excess flux from a background source). 
To test the dependence of the derived mass on this measurement, another MCMC run was tested excluding this datapoint, but the overall fit still preferred high disk masses.  Subsequent to these initial MCMC runs, \citet{2016ApJ...823..160D} published their ALMA 2.8 mm continuum observations, from which they found a total disk mass (gas+dust assuming a gas to dust ratio of 100) of $0.057 \pm 0.002 M_{\odot}$. The excellent agreement between these independent results (their estimate and our result from fits without including their 2.8 mm data point) provides increased confidence in the apparently high mass of this disk. Our final MCMC fit using the covariance framework included the 2.8 mm measurement as well as the other photometry.

A key aspect here is the assumed dust opacity. The mass estimate by \citet{2016ApJ...823..160D} was made under the assumption that the disk is optically thin, in which case the mass may be directly computed via
$M (\mathrm{gas + dust}) = \frac{F_{\nu} D^{2}}{\kappa_{\nu} B_{\nu}(T)}$. The derived mass thus depends on both the opacity and the disk temperature. \citet{2016ApJ...823..160D} assumed a disk average temperature of T=10 K. Our best-fit MCFOST model yields the disk internal temperature as a byproduct of the MCMC radiative transfer calculation, and the results are fairly consistent: a calculated disk midplane temperature of 10 K at 100 AU, increasing inwards to 30 K at 5 AU. The larger source of potential systematic bias in the disk mass is thus the assumed dust opacity.  \citet{2016ApJ...823..160D} use a dust opacity characteristic of coagulated dust grains with thin icy mantles \citep[$\kappa = 0.23$, cm$^2$/g][]{1994A&A...291..943O}. Our model uses olivine dust as described in section \ref{modelintro}, which yields a similar opacity at 2.8 mm within a factor of 2. But other results can easily be obtained. For instance, if we instead adopt the dust opacity law from \citet[][]{1990AJ.....99..924B} ($\kappa_\nu = 0.03$ cm$^2$/g at 870 \micron) and use T=20 K, then that yields an estimated disk mass $M (\mathrm{gas + dust}) = \frac{F_{\nu} D^{2}}{\kappa_{\nu} B_{\nu}(T)} = 0.0055 M_{\odot}$, a factor of ten lower (again assuming the standard gas to dust ratio of 100). Better constraints on the dust particle properties and thus the millimeter opacities could help clarify the true mass of this disk. Some disagreement between predictions from the 2.8 mm and 870 $\mu$m continuum measurements may be unsurprising since derived dust masses from sub-mm data may be biased downwards in the case of EODs if they begin to become optically thick at that wavelength. That said, the good agreement between the \citet{2016ApJ...823..160D} millimeter-continuum-derived mass and the result from our fit to the full SED and the HST scattered light image seems to indicate that the relative importance of absorption/emission and scattering of the dust model (which includes, but is not limited to, the dust albedo) used here is a reasonable approximation.

Lastly, we note that a reduced gas-to-dust ratio would of course directly affect the inferred total gas+dust to star mass ratio, but the available observations do not provide any evidence towards (or against) such a hypothesis.

\subsection{Scale Height}

The best fit scale height of $16.2^{+1.7}_{-2.0}$ au (at 100 au) is 
consistent with the low mass of the central star. For a disk that is pressure supported and vertically isothermal with temperature, the Gaussian vertical density distribution is described by Equation \ref{Eq:scaleheight} (Burrows et al. 1996): 

\begin{equation}
  H(r) \, = \sqrt{ \frac{ k_{B}T(r)r^{3}}{ G M_{\mathrm{star}} \mu m_{p}}} 
  \label{Eq:scaleheight}
\end{equation}

\noindent where we assume a reduced mass ($\mu$) of 2.3 If we adopt the best fit scale height value of 16.18 au at a reference radius of 100 au and calculate the temperature of the disk at this radius, we obtain T $\sim$ 23 K. This disk temperature agrees well with observations of other edge-on disks (e.g. HH 30;  \citealp{1996ApJ...473..437B}). 

Additionally, MCFOST is capable of producing the temperature structure within the disk along with the images and SEDs. This can be used as a cross check on the physical self-consistency of our best fit model parameters. The mass averaged temperature (across the vertical direction) for our best fit model at the reference radius (100 au) is $T = 29$~K. 
Surface effects that are exacerbated in scattered light could account for the slight discrepancy between the analytically and numerically estimated disk temperatures, as the surface gas is super heated by stellar radiation. 
The agreement between the dust scale height inferred from the image and the gas scale height computed from the model suggests that the dust grains are well-mixed vertically with very little dust settling, at least for the small dust particles ($\lesssim$ 10 $\mu$m) that dominate the opacity at visible wavelengths.

\subsection{Flaring Exponent}

The best fit flaring exponent was $\beta = 1.19^{+0.09}_{-0.08}$. 
%For a disk with dust that is well mixed with the gas, the disk is vertically isothermal. 
\citet{1987ApJ...323..714K} provide an analytical model for the temperature profile of a flared disk wherein the surface layers are heated by the direct stellar radiation and the energy is re-radiated thermally. Assuming the gas and dust are well mixed vertically, and that the incident angle of the stellar radiation on the flared surface is small, $T(R) = T(R_{0}) \big( \frac{R}{R_{0}} \big)^{2 \beta - 3}$.
We fit the modeled mass-averaged disk temperature profile to this analytic solution, revealing that a flaring exponent of $\beta = 1.29$ is preferred. This value is consistent within $\sim$ 1$\sigma$ of the model preferred value. 
The best fit value is slightly shallower than has been predicted for other young, flared disks with $\beta = 1.3 - 1.5$ \citep[e.g.][]{1997ApJ...490..368C}. This could be an indication of early dust settling in the disk, decoupling the dust and gas and changing the disk thermal pressure profile. 
In this model, dust particles of all sizes are assumed to be evenly distributed vertically throughout the disk. An investigation into the effect of settling of larger grains to the disk midplane is left for future work.

\subsection{Surface Density Exponent}

The surface density exponent is best fit by $\alpha = -1.77^{+0.94}_{-0.14}$, which is near the lower edge of the allowed parameter space. However, allowing for steeper surface density profiles would push the models into a highly unphysical range. The SED favors a very steep surface density profile \citep[also seen for HV Tau C: ][]{2010ApJ...712..112D}, while the images favor a shallow profile with a more gradual taper at the disk edge. It is possible that the steep best fit surface density profile is a reaction to the large disk masses required to fit the mm data in the SED, whereby mass is being concentrated in the center of the disk, where our dataset is poorly equipped to constrain the disk properties.  The SED was not expected to have a strong dependence on the surface density slope. The disk is presumed to be very optically thick across most of the IR portion of the SED. Consequently, the surface density profile would not impact the location of the disk scattering surface which is intercepting and re-radiating light from the central star. It is possible that a degeneracy between surface density exponent and some other star/disk property is influencing this fit (e.g. stellar luminosity, dust albedo, etc.). 

It is unexpected that a disk surface density power law would be steeper than the $\alpha = - 1.5$ value for the minimum mass solar nebula \citep{1977Ap&SS..51..153W}. 
Indeed, \citet{2007ApJ...659..705A} conducted a resolved submillimeter continuum survey of circumstellar disks and find a mean value of $\alpha = - 0.5$.
Instead this steep profile is probably indicative of some shortcoming in our model parameterization. Invoking separate power laws for the inner and outer regions of the disk may provide a solution, but is beyond the scope of this paper.
While we have spatially resolved images at optical wavelengths, the disk is highly optically thick, causing any effects of radial density gradients to be undetectable. Characterizing these would require resolved images at wavelengths where the disk is optically thin (e.g. resolved millimeter continuum images, though it is uncertain if the disk is truly optically thin at these wavelengths). Scattered light imaging alone simply does not constrain the surface density exponent in the innermost regions of the disk.
Previous studies of the radial structure of protoplanetary disks observed in millimeter continuum find surface density profiles that are generally shallower than presented here, though there are a few exceptions \citep[e.g. DG Tau, GM Aur;][]{2011A&A...529A.105G}. 
Estimates of the surface density distributions inferred from resolved mm data at different wavelengths vary widely \citep{2010ApJ...714.1746I}, suggesting that these disks are not optically thin even in the 1-3 mm range.

\section{Discussion}
\label{Sec:discussion}

\subsection{Mass and Stability of the Disk}

The best fit dust mass ($0.00057 \, M_{\odot}$ or $190 \, M_{\oplus}$) and the associated total (gas + dust) disk mass ($60 \, M_{Jup}$, assuming a gas-to-dust ratio of 100) imply a disk mass to star mass ratio ($M_{D}/M_{\mathrm{star}}$) significantly higher than expected for its age and spectral type. \citet{2011ARA&A..49...67W} provide a review of protoplanetary disks and report a relatively flat distribution of disk masses when spaced logarithmically, with a sharp drop outside of $\sim$ 50 $M_{Jup}$, and an average disk mass to host stellar mass ratio of 0.01 albeit with large scatter. The median mass (assuming a gas-to-dust ratio of 100) of disks around GKM spectral type hosts is 5 $M_{Jup}$ (implying a dust mass of $\sim$16 $M_{\Earth}$).

\begin{figure}[hbpt!]
\begin{center}
\includegraphics[width=3.0in]{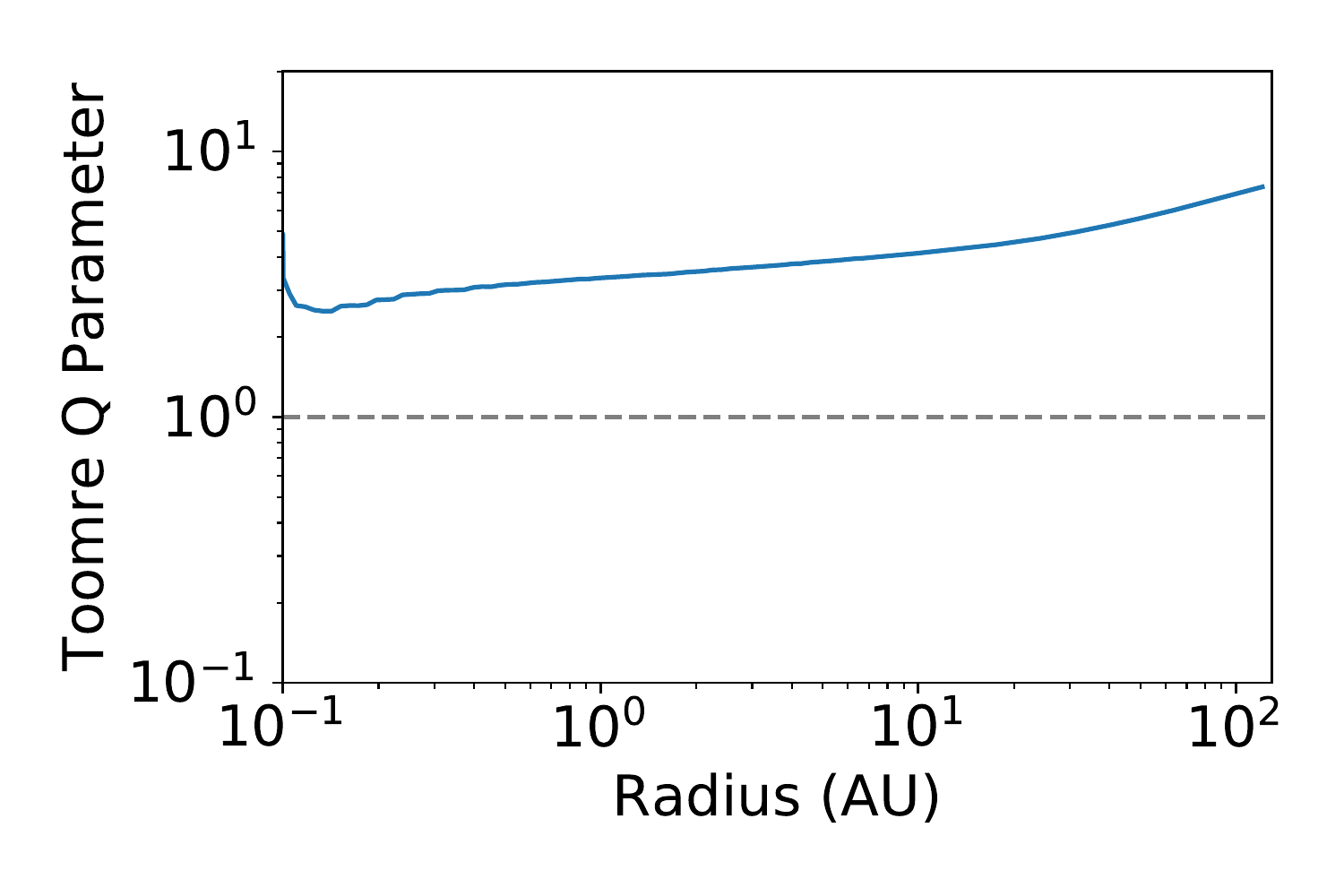}
\caption{Radial profile of the Toomre Q parameter for our best fit disk. The disk appears to be stable at all radii.}
\label{fig:Q}
\end{center}
\end{figure}

This trend of low $M_{D}/M_{\mathrm{star}}$ mass ratios seems to continue for low mass stars.  \citet{2016ApJ...819..102V} conducted a survey of disk masses for low mass stars with ALMA, finding a range of masses between 0.1 and 1 $M_{\Earth}$ for their eight targets. One target in their sample, Allers 8 (an M3 star with a mass of 0.34 $M_{\odot}$), has similar stellar parameters to ESO~H$\alpha$~569, but a significantly lower dust mass of 1.05 $M_{\Earth}$. However, the bulk of the disks in their sample are located in the Upper Scorpius SFR \citep[$\sim$ 10 Myr;][]{2016ApJ...816...21D} and are older than our target. The authors have an additional dataset for the younger Taurus SFR, with preliminary estimates for the dust mass upper limit of 25 $M_{\Earth}$ for a sample of stars with an earliest spectral type of M4 (Ward-Duong, private communication). 
Additionally, \citet{2013ApJ...771..129A} conduct a survey of the protoplanetary disks with low mass hosts (spectral types earlier than M8.5) in the Taurus SFR and find slightly higher disk masses. The authors estimate the disk masses from their mm-wave continuum luminosity, and find that the median disk mass to stellar mass ratio is 0.3\%, with very few disks having a ratio of $\ge$ 10\%. Targets in their sample in the M3-M4 spectral type range have disk dust masses of 2-17 $M_{\Earth}$, with an average of 9 $M_{\Earth}$.

While uncommon, protoplanetary disks with large disk masses aren't unprecedented. \citet{2010ApJ...712..112D} model scattered light images and SEDs for the HV Tau C system and find a best fit dust mass of $M_{dust} \ge 10^{-3} \, M_{\odot}$ which gives $M_{D}/M_{\mathrm{star}} \sim 0.2$ (meaning the disk is 20\% the mass of the central star) assuming a gas-to-dust ratio of 100. 
Likewise, \citet{2003A&A...400..559D} model a mm image of the HK Tau B protoplanetary disk and get a best fit total disk mass of $M_{disk} \simeq 2 \times 10^{-2} \, M_{\odot}$, which gives $M_{D}/M_{\mathrm{star}} \sim 0.04$ (4\% of the stellar mass). 
\citet{2008A&A...485..531G} present an in-depth study of the IRAS 04158+2805 disk using images in the optical, NIR, polarization maps in the optical and mid-IR and X-ray spectra. The dust mass is constrained to be $M_{dust} = 1.0 - 1.75 \times 10^{-4} \, M_{\odot}$, which also gives $M_{D}/M_{\mathrm{star}} = \sim 0.04$ (4\%). All three disks are in the Taurus SFR, and the first two disks above are in multiple systems. Likewise, all of these sources are viewed edge-on. It is possible that the large inferred disk masses could be the result of a selection effect (observations of edge-on disks are only sensitive to the most massive disks), or some artifact of our fitting method which compensates for missing physics by placing more mass in the disk. 
The fit for the dust mass is driven by the SED, but the spectral coverage is poor in the millimeter. The mass estimates could be reduced by including larger opacities in the mm, for instance by adding amorphous carbon into the mixture, or by using a more complex, nonuniform particle distribution.

Throughout this paper, we have assumed a gas to dust mass ratio of 100, as is typical of other young disks and the ISM. However, very recent work by \citet{2017arXiv170603320L} estimate the gas mass around ESO H$\alpha$ 569 from ALMA $^{13}$CO line emission and find only $\sim 1.3 \, M_{Jup}$ of gas mass in the disk, though optical depth effects and details of the CO freeze-out are likely to introduce major sources of uncertainty. Combined with our own dust mass estimate, this gives an uncharacteristically low gas to dust ratio of only $\sim 2$. 
While gas depletion in the disk would lower the unusually high best fit total disk mass, the flared appearance strongly confirms this is a young pressure-supported gas+dust disk. A gas to dust ratio of 2:1 is suggestive of a later evolutionary stage. 
This disagreement highlights the challenges of measuring disk masses for EODs, which are generally optically thick even at millimeter wavelengths.

The MCMC radiative transfer fit prefers a disk with an abnormally large disk mass that is $\sim$ 16\% the mass of the central star. 
We investigate the stability of the disk via the Toomre Q parameter. 

\begin{equation}
  Q = \frac{c_{s} \kappa}{\pi G \Sigma}
\end{equation}

\noindent where $c_{s}$ is the sound speed in the disk, $\kappa$ is the epicyclic frequency, and $\Sigma$ is the surface density profile of the disk. For a vertically isothermal disk with a Keplerian velocity, $\kappa = \Omega = \sqrt{\frac{G M_{\mathrm{Star}}}{R^{3}}}$ and a sound speed $c_{s} = \sqrt{\frac{k_{B} T}{\mu m_{p}}}$ where a reduced mass ($\mu$) of 2.3 was assumed. Figure \ref{fig:Q} shows the radial profile of the Toomre Q parameter. It shows that the disk appears to be stable at all radii. 

It is worth noting here that a good constraint is not expected on the properties of the inner regions of the disk from scattered light imaging and the SED alone. Any change in the interior structure (e.g. an inner wall, spiral structure, or a broken surface density power law) of the disk would affect stability. 
Each of these mechanisms would increase the variability of the system, possibly accounting for the observed variability in several of the photometric points included in the SED.

\subsection{ESO~H$\alpha$~569 Compared to Other Cha I Disks}

\citet{2014MNRAS.443.1587R} conducted a survey of disks in Cha I as identified from IR excesses in the SEDs. For 34 objects, disk masses were estimated. The median of the distribution of disk masses is 0.005 $M_{\odot}$, which corresponds to 0.5\% of the stellar mass, while the tail of the distribution stretches to 0.1 $M_{\odot}$ for more massive central stars. ESO~H$\alpha$~569 is a clear outlier with 10 times more mass than the median value.

The \citet{2007ApJS..173..104L} survey of Cha I names six members as likely edge-on disk candidates because they are underluminous for their spectral type and are seen in scattered light (CHSM 15991, T14A, ISO 225, ESO~H$\alpha$~569 and 574, and Cha J11081938-7731522).
One of those objects, Cha J11081938-7731522, appears extended in their survey with a butterfly morphology, providing further support that these targets are all likely edge-on disks. 
Two members of that list, ESO~H$\alpha$~569 and 574, were observed in our HST campaign, which confirmed that both are edge-on protoplanetary disks.

\subsection{A Deficit of Edge-on Disks?}
\citet{2008ApJ...675.1375L} use Spitzer colors to estimate the disk fraction as a function of stellar mass. For stars of spectral type between K6 and M3.5 the disk fraction in Cha I is 0.64 $\pm$ 0.06 disks per star. If we multiply this fraction by the fraction of disks expected to have inclinations between 75 and 90 degrees, roughly 17\% of stars with young disks should host edge-on disks. 
However, a recent survey of 44 YSOs hosting circumstellar disks detectable with Herschel found only 2 edge-on disks \citep{2014MNRAS.443.1587R} as classified from the SEDs.
While the sample size of this survey is small, 
this surprising lack of known EODs is a common phenomena seen for many SFRs \citep{2014IAUS..299...99S} and was one of the key motivating factors for our HST survey. While that program doubled the number of known EODs, the increased sample remains smaller than would be predicted from purely geometrical grounds. This suggests that many disks must be near-edge-on but with insufficient material and/or vertical extent to block the direct light of the star. 
Flatter disks, with lower H/R values than ESO~H$\alpha$~569, would only appear edge-on for a narrower inclination range. For instance if the ``typical'' young disk is flared enough to only occult its star within 5 degrees, Considering a range from 85 to 90 degrees would give an edge-on disk fraction per star of 4\%, more in line with what is observed.

\begin{figure*}[hbpt!]
\begin{center}
\includegraphics[width=6.5in]{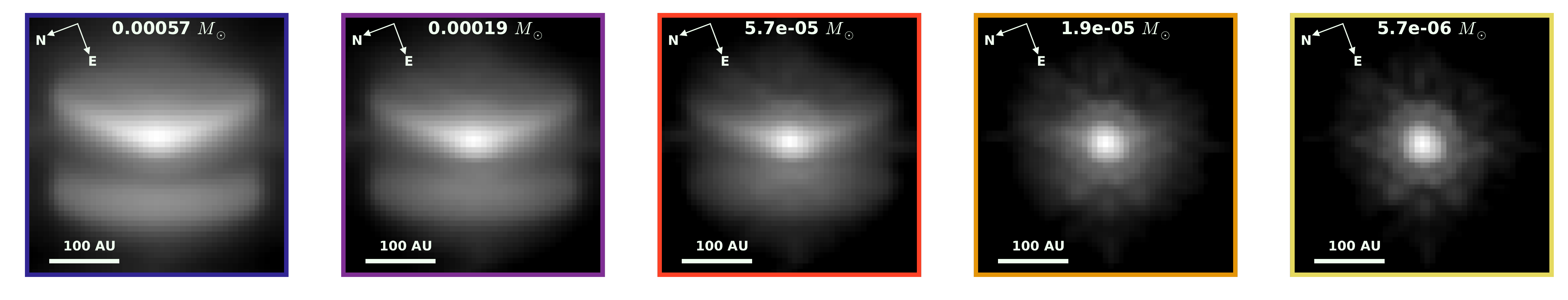}
\includegraphics[width=4.5in]{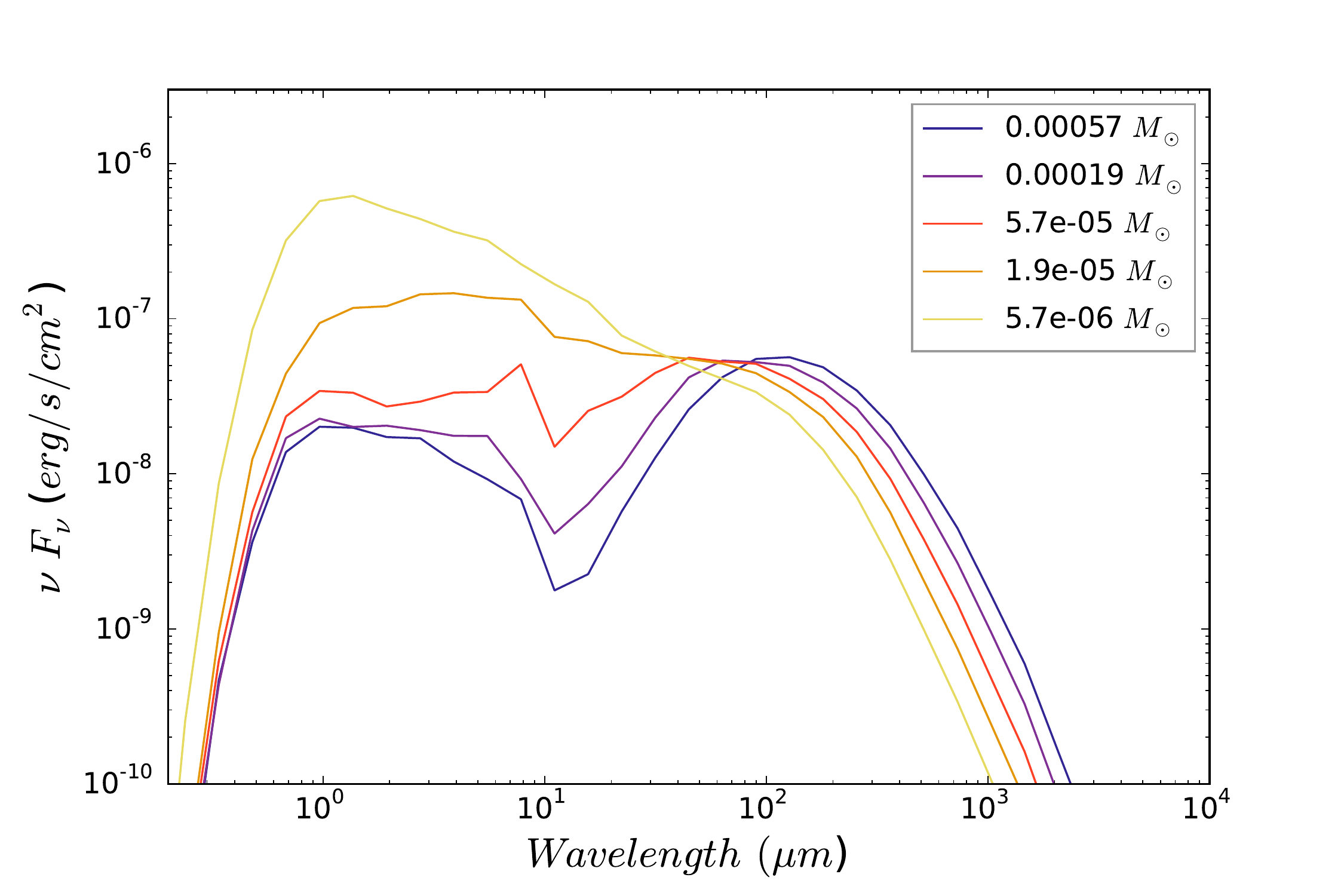}
\caption{We show the evolution of the shape of the image and SED for different dust masses. Our best fit model is shown in blue. The other models use the same parameter values except for the mass, which is some fraction of the best fit dust mass as indicated in the legend. For a fixed inclination, decreasing the dust mass moves photons from the thermal peak in the SED to the scattered light peak. Decreasing the mass by a factor of 10 generates a flat SED without the double peaked structure. Likewise, if the dust mass is one tenth the best fit value, the double nebula shape begins to disappear in the scattered light image, and is not seen at all in the $1.9 \times 10^{-5} \, M_{\odot}$ model. \label{fig:mass}}
\end{center}
\end{figure*}

Alternatively, this could suggest that the 'typical' double peaked SED assumed for edge-on disks may only present for the disks with an unusually high disk mass. 
The targets for this edge-on disk survey were selected based on the shape of the SEDs. Specifically, targets with a doubled peaked SED, where the stellar peak flux was of order the same as the dust peak flux in the IR. Figure \ref{fig:mass} shows the effect of changing dust mass on the structure of the SED, and the scattered light image for a fixed inclination. Disk masses shown are for the best fit disk mass divided by factors of 3, 10, 30 and 100. After dividing by a factor of 10 (for a more reasonable $M_{D} / M_{\odot} \sim 0.016$) the double peaked structure has disappeared, and we would not have included this target in our sample. It is possible that this could account for the relative lack of known edge-on disks; the selection metrics used are biased towards detecting only the most massive disks, as they require fairly large line-of-sight opacities. 

It is therefore possible that the edge-on disk detections thus far are outliers in the population of young disks. Double-peaked SEDs alone are an insufficient indicator of the edge-on disk fraction, and images in scattered light or thermal emission with high spatial resolution are required to determine the true nature of these objects. Existing surveys of young, nearby SFRs tend to have selection biases towards more face-on systems and are dependent on the cloud properties and the science drivers of the survey. In order to determine the true edge-on disk fraction and to confirm or deny that the high disk mass of ESO~H$\alpha$~569 is indeed representative of the population of protoplanetary disks, a uniform sample of disk observations at sufficiently high angular resolution will be required.

\section{Summary and Conclusions}
\label{Sec:summary}

We have resolved the disk around ESO~H$\alpha$~569 in scattered light with HST/ACS and unambiguiously confirmed that it is an optically thick protoplanetary disk viewed nearly edge-on. We performed radiative transfer modeling using a variety of fitting techniques to constrain the geometry and grain properties of the disk. 
We successfully combine a covariance-based log likelihood estimation with an MCMC framework to simultaneously fit the scattered light image and literature compiled SED for ESO~H$\alpha$~569. 
Our main results are as follows:

\begin{itemize}

\item[--] We find that a tapered-edge disk structure, with an exponential falloff of material outside of the apparent outer radius, is necessary to generate the diffuse scattered light emission above the disk midplane, the flux ratio between the top/bottom nebulae of the disk, and the width of the dark-lane simultaneously.

\item[--] The best fit disk mass of 0.057 $M_{\odot}$ is abnormally large, especially considering the small central object, though multiple mm continuum observations support this estimate. Assuming a gas to dust ratio of 100, the disk mass is 16\% the mass of the central star, establishing the ESO H$\alpha$ 569 disk as a clear outlier in the Cha I SFR. 
Despite the high mass, the disk appears gravitationally stable at all radii.

\item[--] The vertical structure of the disk as defined by the scale height and the power law flaring exponent is well constrained. The best fit model has a mass-averaged disk temperature of $\sim$ 23 K, similar to other disk observations. The scale height is self-consistent with the modeled temperature profile, supporting a flared disk model in which the gas and dust are well-coupled.

\end{itemize}

A large effort was put into simultaneous and consistent fitting of the images and the SEDs, resulting in a disk model that is is a good compromise between the two.  But naturally a separate fit to each individual observable is capable of yielding a better fit to that one, at the cost of an inferior fit to the other. This is likely due to: (1) Limitations in the parameterization of the complex physical processes ongoing in protoplanetary disks. In this work, a fairly simple analytic disk structure formalism with a single grain population was used. (2) The inability of our dataset to constrain some aspects of  relevant physics and processes (e.g. neither the SED nor the scattered light image provide much information on the innermost regions of the disk).

Using a combination of different observables (spectral data, images in scattered light, and thermal emission, and polarimetry data to constrain grain properties) helps to break degeneracies between various model parameters. However, care must be taken to determine the correct approach for the relative weighting of observables with different noise properties and model sensitivities. Now that high contrast imaging systems designed to study these circumstellar environments in greater detail are coming on line, there is a plethora of great observations for disks in a wide range of evolutionary stages which formed under a range of initial conditions. We may be entering an era where we have statistically significant numbers of circumstellar disk observations to employ population synthesis techniques. This is an important step if we hope to understand the inherent physics in the disk and planet formation processes. 
The tools we have been developing take us a step closer to being able to consistently make fits and measurements to e.g. the entire known sample of edge-on disks. 

To better constrain the ESO~H$\alpha$~569 disk and stellar parameters, we would need to incorporate resolved images at multiple wavelengths. In this work we chose not to model the 0.6 $\mu$m image because of the contamination from the jet. However, we have recently obtained resolved images in the HST F475W filter and will use this to probe the diffuse scattering material high up above the disk in a forthcoming paper. Additionally, an ALMA Cycle 4 program (PI: F. M{\'e}nard) was awarded to map the thermal emission from 15 confirmed edge-on disks from our HST sample at 870 $\mu m$ and 2 mm to probe dust settling, migration and grain growth. Spatially resolved millimeter observations should go a long way toward disentangling many of the outstanding uncertainties regarding this disk's structure. 
%and b) map the CO emission from 2 select targets to study the (de)coupling of gas and dust. 
Looking forward, with the launch of JWST, the MIRI MRS integral field spectrograph will provide spatially- and spectrally-resolved data across the entire 5 - 30 $\mu$m range for many disks. This would not only help to constrain the structure of the disk in a regime where the current SED fit particularly struggles, but also provide valuable and detailed information about the dust species within the disk.

\acknowledgements
The authors thank K. Ward-Duong for helpful discussion.
This material is based upon work supported by the National Science Foundation Graduate Research Fellowship under Grant No. DGE-1232825. Based on observations made with the NASA/ESA Hubble Space Telescope, obtained at the Space Telescope Science Institute, which is operated by the Association of Universities for Research in Astronomy, Inc., under NASA contract NAS 5-26555. These observations are associated with program \#12514. C. Pinte acknowledges funding from the European Commission's 7$^\mathrm{th}$ Framework Program (contract PERG06-GA-2009-256513) and from Agence Nationale pour la Recherche (ANR) of France under contract ANR-2010-JCJC-0504-01. F. M\'enard acknowledges funding from ANR of France under contract number ANR-16-CE31-0013. 
 
\software{MCFOST \citep{2006A&A...459..797P}, Tiny Tim \citep{1995ASPC...77..349K}, emcee \citep{2013PASP..125..306F} mcfost-python (https://github.com/swolff9/mcfost-python)}

\appendix

Here we provide a simplified disk model fitting effort designed to illustrate the effect of the two `goodness of fit' metrics used in the MCMC explorations of parameter space: $\chi^{2}$ and covariance log-likelihood based estimation described in Sections 3.4.1 and 3.4.2 respectively. While this demonstrates the power of the two tools, we recognize that it is not a comprehensive test of performance. A full benchmarking effort of the \texttt{mcfost-python} package is beyond the scope of this paper. 

To test the ability of both fitting metrics, we generate an MCFOST model with known parameter values, add randomly generated 1$\sigma$ noise to both the MCFOST produced image and SED, and attempt to retrieve the parameters. The model was randomly drawn from the ESO~H$\alpha$~569 MCMC chain described above. We perform a fit to this synthetic dataset using both the $\chi^{2}$ and covariance log-likelihood based estimation. For simplicity, we choose only to fit the scale height and inclination of our modeled disk. We expect both methods to recover the known parameter values within the uncertainties. 
Parameter values used for the synthetic dataset are shown in Table \ref{table:syntheticdata}.

To illustrate the power of the covariance framework over the $\chi^{2}$ fitting technique, we perform the same test, but purposefully input a disk dust mass too low by a factor of 10 into the MCFOST parameter file. This will test how robust the covariance framework in the presence of clear limitations in the model's ability to fit the data. 

In an effort to conduct these tests as close to the MCMC results reported above, we use the same Parallel Tempered ensemble sampler with two temperatures and 50 walkers. With only two free parameters, the chains converged more quickly, requiring only $N_{\mathrm{steps}} = 10000$ with $N_{\mathrm{burn}} = 0.2 \, N_{\mathrm{steps}}$. The allowable parameter ranges for the inclination and scale height were the same as reported in Table 2.

\begin{figure}[h!]
\begin{center}
\includegraphics[width=3.2in]{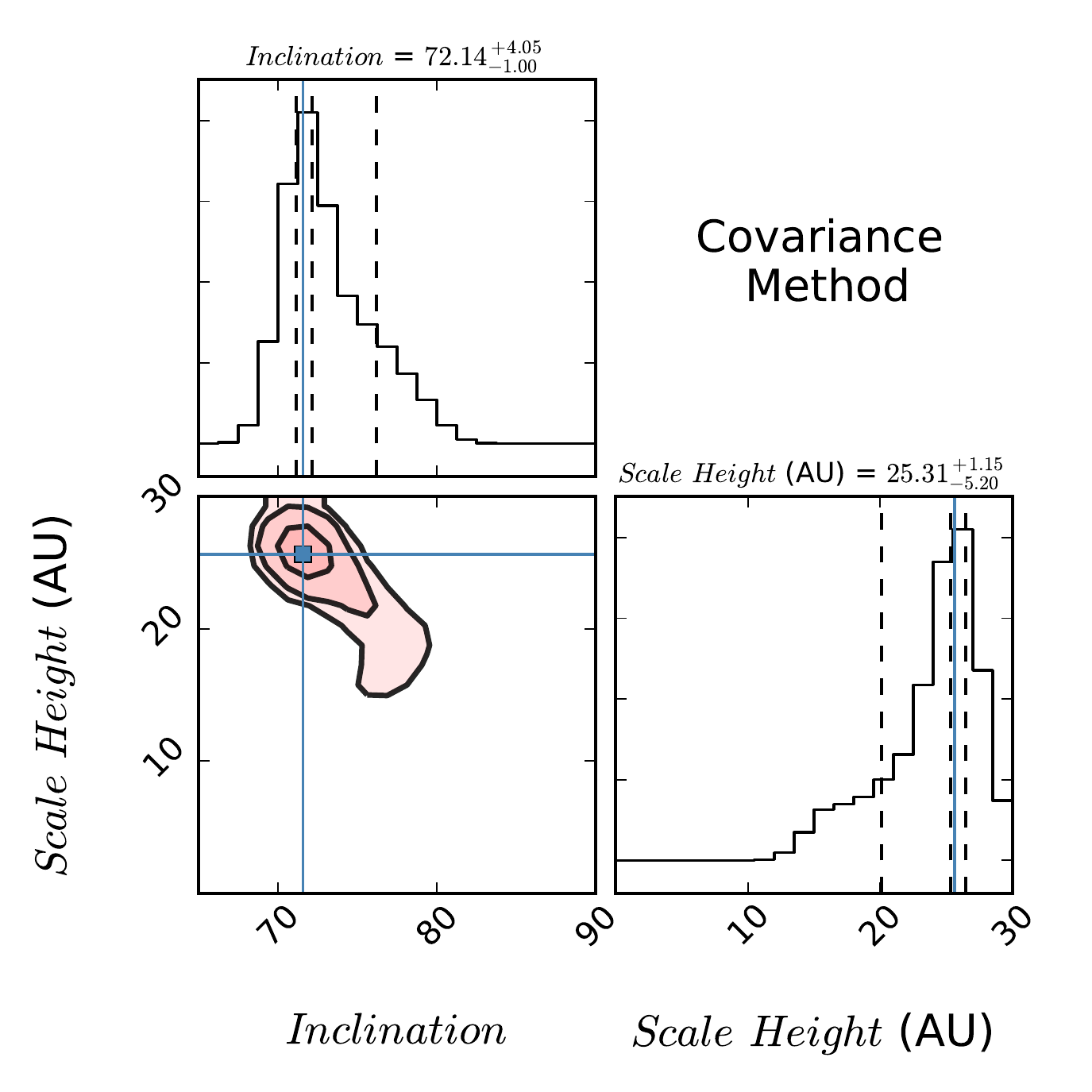}
\includegraphics[width=3.2in]{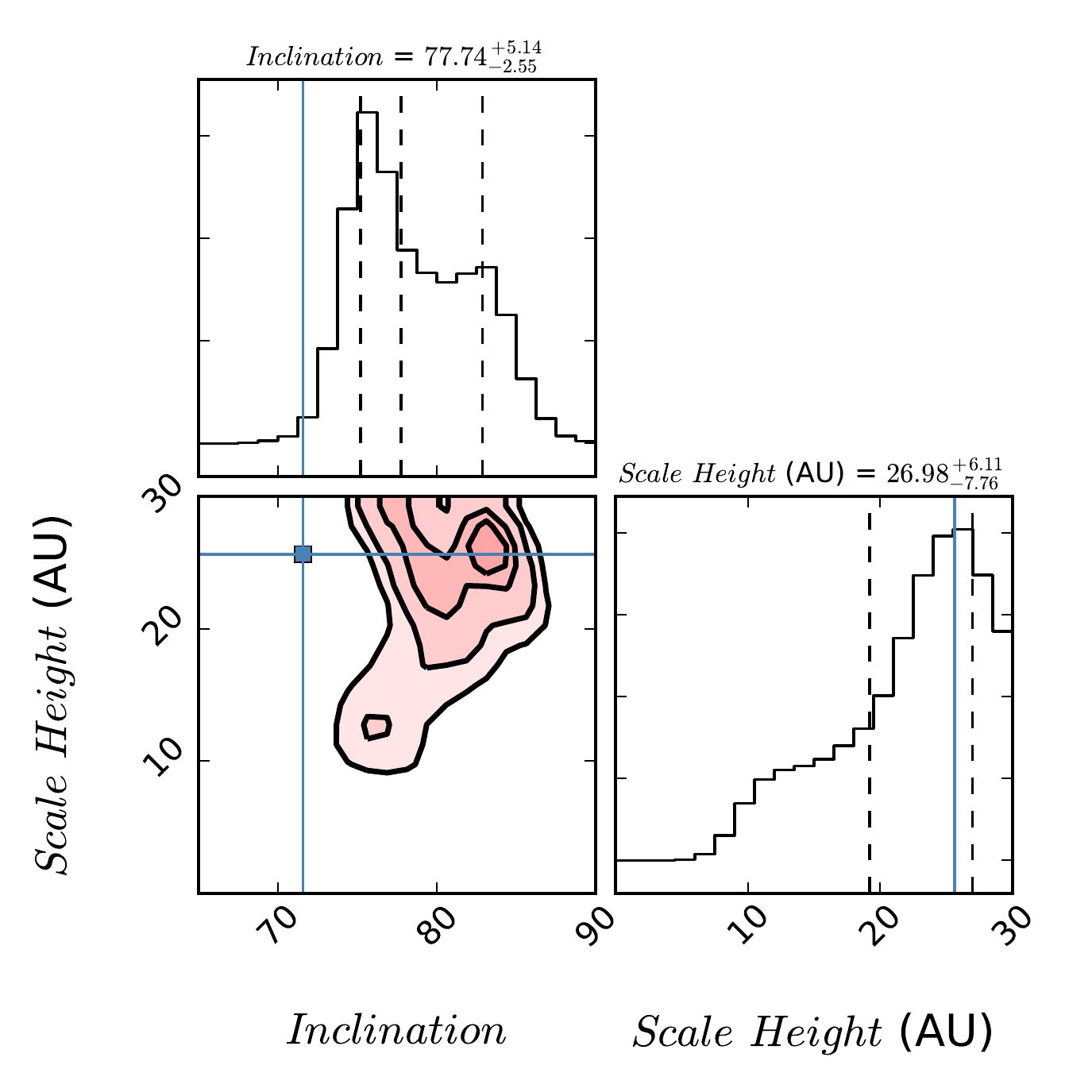}
\caption{\textbf{Left:} MCMC results of the covariance log-likelihood estimation fit using a synthetic dataset. We fit only the scale height and inclination of the modeled disk. The blue lines correspond to the known values for each parameter. The correct parameter values were retrieved, and the distributions are sharply peaked. \textbf{Right:} Same as the left panel, but the MCMC run was conducted using an incorrect disk dust mass in the MCFOST parameter files. Even assuming a depleted dust mass, the scale height of the disk is still recovered, while the best fit inclination is $\sim \, 2\sigma$ discrepant. The covariance framework is less sensitive to any global limitations of the disk model to fit the given dataset. \label{fig:covar}}
\end{center}
\end{figure}

\begin{figure}[h!]
\begin{center}
\includegraphics[width=3.2in]{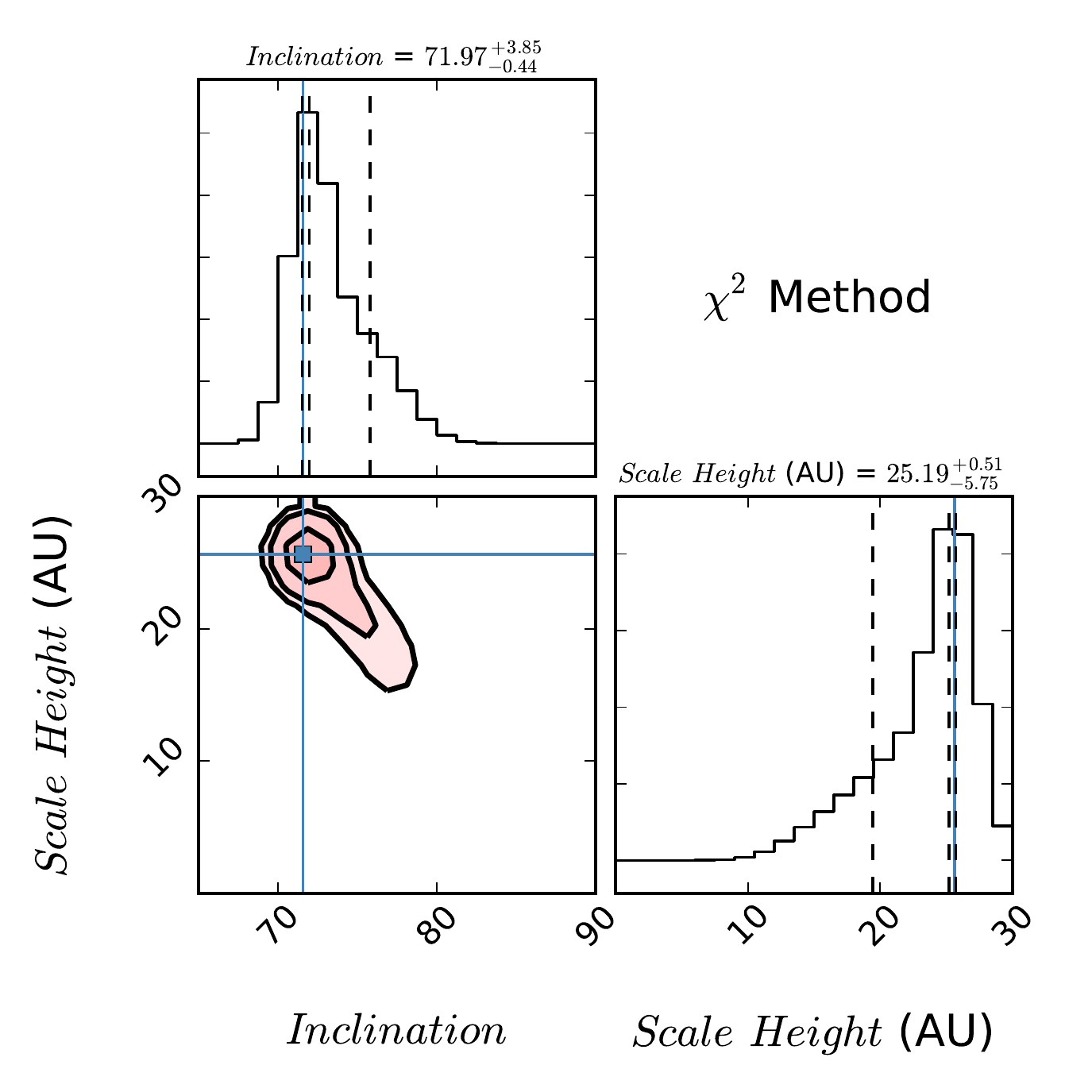}
\includegraphics[width=3.2in]{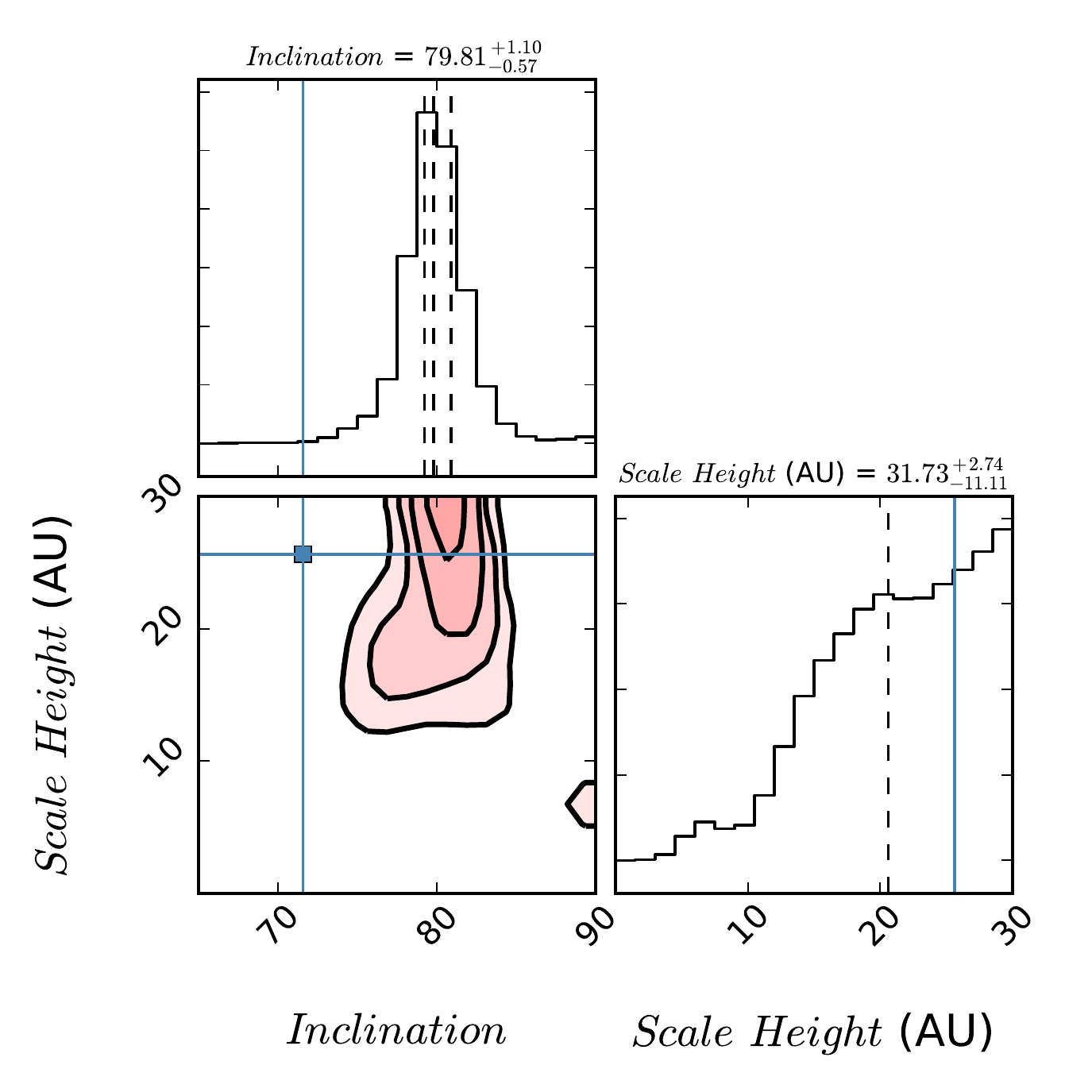}
\caption{ Same as figure \ref{fig:chisq} but the MCMC was run using the $\chi^{2}$ log-likelihood based estimation rather than the covariance framework. Unlike the covariance case, the $\chi^{2}$ fitting metric has a difficult time retrieving any of the correct parameter values when the incorrect disk dust mass was used to generate each MCFOST model. The disk scale height is not well constrained at all, and the best fit inclination is $\sim \, 14\sigma$ discrepant.  \label{fig:chisq}}
\end{center}
\end{figure}

Figures \ref{fig:covar} and \ref{fig:chisq} present the results for the Covariance and $\chi^{2}$ fitting techniques, respectively. Both methods retrieve the input inclination and scale height within the uncertainties when using the correct dust mass. However, when the dust mass is set to one tenth the actual value, both fitting methods struggle to retrieve the correct parameter values. The covariance run successfully recovered the disk scale height, though the uncertainties are larger than the correct dust mass case. The inclination was found to be $77.7^{+5.1}_{-2.6}$ degrees, which is only $\sim 2\sigma$ discrepant from the true value. With the incorrect disk mass, the $\chi^{2}$ run was unable to recover either parameter. The scale height of the disk is not well constrained at all, while the likelihood distribution for the inclination is sharply peaked at $79.8^{+1.1}_{-0.6}$ degrees, which is $\sim  14\sigma$ discrepant from the true value. It is unsurprising that the covariance framework is much more robust to global limitations of the models to fit the dataset.

 \begin{deluxetable}{lcl}[htpb!]  % <--- column justification (center/left/right)
\tablecolumns{3}
\tablecaption{Parameter values for the Synthetic Dataset}
\tablehead{   % column headings
  \colhead{Parameters} &
  \colhead{Values} &
  \colhead{Notes}
 % \colhead{Total}
}
\tablewidth{5.0in}
\startdata
Inclination & $71.6^{\circ}$ & Allowed to vary. \\
Scale Height (R=100 au)  & 25.6 au & Allowed to vary. \\ 
Dust Mass  &  $4.94 \times 10^{-4} \, M_{\odot}$ & Held constant\tablenotemark{a} \\ 
Surface Density $\alpha$ & -1.76 & Held constant. \\
Flaring $\beta$     & 1.54 & Held constant. \\
\enddata
\tablenotetext{a}{This value was held constant for all runs, however, a value of $4.94 \times 10^{-5} \, M_{\odot}$ (0.1 times the actual value) was used to test the robustness of the fitting techniques to systematic model errors. }
\label{table:syntheticdata}
\end{deluxetable}

\bibliography{main}

\begin{thebibliography}{}
\expandafter\ifx\csname natexlab\endcsname\relax\def\natexlab#1{#1}\fi
\providecommand{\url}[1]{\href{#1}{#1}}

\bibitem[{{Andrews} {et~al.}(2013){Andrews}, {Rosenfeld}, {Kraus}, \&
  {Wilner}}]{2013ApJ...771..129A}
{Andrews}, S.~M., {Rosenfeld}, K.~A., {Kraus}, A.~L., \& {Wilner}, D.~J. 2013,
  \apj, 771, 129

\bibitem[{{Andrews} \& {Williams}(2007)}]{2007ApJ...659..705A}
{Andrews}, S.~M., \& {Williams}, J.~P. 2007, \apj, 659, 705

\bibitem[{{Bally} {et~al.}(2006){Bally}, {Walawender}, {Luhman}, \&
  {Fazio}}]{2006AJ....132.1923B}
{Bally}, J., {Walawender}, J., {Luhman}, K.~L., \& {Fazio}, G. 2006, \aj, 132,
  1923

\bibitem[{{Baraffe} {et~al.}(1998){Baraffe}, {Chabrier}, {Allard}, \&
  {Hauschildt}}]{1998A&A...337..403B}
{Baraffe}, I., {Chabrier}, G., {Allard}, F., \& {Hauschildt}, P.~H. 1998, \aap,
  337, 403

\bibitem[{{Beckwith} {et~al.}(1990){Beckwith}, {Sargent}, {Chini}, \&
  {Guesten}}]{1990AJ.....99..924B}
{Beckwith}, S.~V.~W., {Sargent}, A.~I., {Chini}, R.~S., \& {Guesten}, R. 1990,
  \aj, 99, 924

\bibitem[{{Belloche} {et~al.}(2011){Belloche}, {Schuller}, {Parise},
  {Andr{\'e}}, {Hatchell}, {J{\o}rgensen}, {Bontemps}, {Wei{\ss}}, {Menten}, \&
  {Muders}}]{2011A&A...527A.145B}
{Belloche}, A., {Schuller}, F., {Parise}, B., {et~al.} 2011, \aap, 527, A145

\bibitem[{{Bertout} {et~al.}(1999){Bertout}, {Robichon}, \&
  {Arenou}}]{1999A&A...352..574B}
{Bertout}, C., {Robichon}, N., \& {Arenou}, F. 1999, \aap, 352, 574

\bibitem[{{Boulanger} {et~al.}(1998){Boulanger}, {Bronfman}, {Dame}, \&
  {Thaddeus}}]{1998A&A...332..273B}
{Boulanger}, F., {Bronfman}, L., {Dame}, T.~M., \& {Thaddeus}, P. 1998, \aap,
  332, 273

\bibitem[{{Burrows} {et~al.}(1996){Burrows}, {Stapelfeldt}, {Watson}, {Krist},
  {Ballester}, {Clarke}, {Crisp}, {Gallagher}, {Griffiths}, {Hester},
  {Hoessel}, {Holtzman}, {Mould}, {Scowen}, {Trauger}, \&
  {Westphal}}]{1996ApJ...473..437B}
{Burrows}, C.~J., {Stapelfeldt}, K.~R., {Watson}, A.~M., {et~al.} 1996, \apj,
  473, 437

\bibitem[{{Cambresy} {et~al.}(1997){Cambresy}, {Epchtein}, {Copet}, {de Batz},
  {Kimeswenger}, {Le Bertre}, {Rouan}, \& {Tiphene}}]{1997A&A...324L...5C}
{Cambresy}, L., {Epchtein}, N., {Copet}, E., {et~al.} 1997, \aap, 324, L5

\bibitem[{{Carmona} {et~al.}(2014){Carmona}, {Pinte}, {Thi}, {Benisty},
  {M{\'e}nard}, {Grady}, {Kamp}, {Woitke}, {Olofsson}, {Roberge}, {Brittain},
  {Duch{\^e}ne}, {Meeus}, {Martin-Za{\"i}di}, {Dent}, {Le Bouquin}, \&
  {Berger}}]{2014A&A...567A..51C}
{Carmona}, A., {Pinte}, C., {Thi}, W.~F., {et~al.} 2014, \aap, 567, A51

\bibitem[{{Chabrier} {et~al.}(2000){Chabrier}, {Baraffe}, {Allard}, \&
  {Hauschildt}}]{2000ApJ...542..464C}
{Chabrier}, G., {Baraffe}, I., {Allard}, F., \& {Hauschildt}, P. 2000, \apj,
  542, 464

\bibitem[{{Chiang} \& {Goldreich}(1997)}]{1997ApJ...490..368C}
{Chiang}, E.~I., \& {Goldreich}, P. 1997, \apj, 490, 368

\bibitem[{{Chiang} {et~al.}(2001){Chiang}, {Joung}, {Creech-Eakman}, {Qi},
  {Kessler}, {Blake}, \& {van Dishoeck}}]{2001ApJ...547.1077C}
{Chiang}, E.~I., {Joung}, M.~K., {Creech-Eakman}, M.~J., {et~al.} 2001, \apj,
  547, 1077

\bibitem[{{Chiang} {et~al.}(2012){Chiang}, {Looney}, \&
  {Tobin}}]{2012ApJ...756..168C}
{Chiang}, H.-F., {Looney}, L.~W., \& {Tobin}, J.~J. 2012, \apj, 756, 168

\bibitem[{{Cleeves} {et~al.}(2016){Cleeves}, {{\"O}berg}, {Wilner}, {Huang},
  {Loomis}, {Andrews}, \& {Czekala}}]{2016ApJ...832..110C}
{Cleeves}, L.~I., {{\"O}berg}, K.~I., {Wilner}, D.~J., {et~al.} 2016, \apj,
  832, 110

\bibitem[{{Comer{\'o}n} {et~al.}(2004){Comer{\'o}n}, {Reipurth}, {Henry}, \&
  {Fern{\'a}ndez}}]{2004A&A...417..583C}
{Comer{\'o}n}, F., {Reipurth}, B., {Henry}, A., \& {Fern{\'a}ndez}, M. 2004,
  \aap, 417, 583

\bibitem[{{Cotera} {et~al.}(2001){Cotera}, {Whitney}, {Young}, {Wolff}, {Wood},
  {Povich}, {Schneider}, {Rieke}, \& {Thompson}}]{2001ApJ...556..958C}
{Cotera}, A.~S., {Whitney}, B.~A., {Young}, E., {et~al.} 2001, \apj, 556, 958

\bibitem[{{Czekala} {et~al.}(2015){Czekala}, {Andrews}, {Mandel}, {Hogg}, \&
  {Green}}]{2015ApJ...812..128C}
{Czekala}, I., {Andrews}, S.~M., {Mandel}, K.~S., {Hogg}, D.~W., \& {Green},
  G.~M. 2015, \apj, 812, 128

\bibitem[{{David} {et~al.}(2016){David}, {Hillenbrand}, {Cody}, {Carpenter}, \&
  {Howard}}]{2016ApJ...816...21D}
{David}, T.~J., {Hillenbrand}, L.~A., {Cody}, A.~M., {Carpenter}, J.~M., \&
  {Howard}, A.~W. 2016, \apj, 816, 21

\bibitem[{{Dorschner} {et~al.}(1995){Dorschner}, {Begemann}, {Henning},
  {Jaeger}, \& {Mutschke}}]{1995A&A...300..503D}
{Dorschner}, J., {Begemann}, B., {Henning}, T., {Jaeger}, C., \& {Mutschke}, H.
  1995, \aap, 300, 503

\bibitem[{{Duch{\^e}ne} {et~al.}(2003){Duch{\^e}ne}, {M{\'e}nard},
  {Stapelfeldt}, \& {Duvert}}]{2003A&A...400..559D}
{Duch{\^e}ne}, G., {M{\'e}nard}, F., {Stapelfeldt}, K., \& {Duvert}, G. 2003,
  \aap, 400, 559

\bibitem[{{Duch{\^e}ne} {et~al.}(2010){Duch{\^e}ne}, {McCabe}, {Pinte},
  {Stapelfeldt}, {M{\'e}nard}, {Duvert}, {Ghez}, {Maness}, {Bouy}, {Barrado y
  Navascu{\'e}s}, {Morales-Calder{\'o}n}, {Wolf}, {Padgett}, {Brooke}, \&
  {Noriega-Crespo}}]{2010ApJ...712..112D}
{Duch{\^e}ne}, G., {McCabe}, C., {Pinte}, C., {et~al.} 2010, \apj, 712, 112

\bibitem[{{Dunham} {et~al.}(2016){Dunham}, {Offner}, {Pineda}, {Bourke},
  {Tobin}, {Arce}, {Chen}, {Di Francesco}, {Johnstone}, {Lee}, {Myers},
  {Price}, {Sadavoy}, \& {Schnee}}]{2016ApJ...823..160D}
{Dunham}, M.~M., {Offner}, S.~S.~R., {Pineda}, J.~E., {et~al.} 2016, \apj, 823,
  160

\bibitem[{{Espaillat} {et~al.}(2011){Espaillat}, {Furlan}, {D'Alessio},
  {Sargent}, {Nagel}, {Calvet}, {Watson}, \& {Muzerolle}}]{2011ApJ...728...49E}
{Espaillat}, C., {Furlan}, E., {D'Alessio}, P., {et~al.} 2011, \apj, 728, 49

\bibitem[{{Flaherty} {et~al.}(2012){Flaherty}, {Muzerolle}, {Rieke},
  {Gutermuth}, {Balog}, {Herbst}, {Megeath}, \& {Kun}}]{2012ApJ...748...71F}
{Flaherty}, K.~M., {Muzerolle}, J., {Rieke}, G., {et~al.} 2012, \apj, 748, 71

\bibitem[{{Foreman-Mackey} {et~al.}(2013){Foreman-Mackey}, {Hogg}, {Lang}, \&
  {Goodman}}]{2013PASP..125..306F}
{Foreman-Mackey}, D., {Hogg}, D.~W., {Lang}, D., \& {Goodman}, J. 2013, \pasp,
  125, 306

\bibitem[{{Glauser} {et~al.}(2008){Glauser}, {M{\'e}nard}, {Pinte},
  {Duch{\^e}ne}, {G{\"u}del}, {Monin}, \& {Padgett}}]{2008A&A...485..531G}
{Glauser}, A.~M., {M{\'e}nard}, F., {Pinte}, C., {et~al.} 2008, \aap, 485, 531

\bibitem[{{Goodman} \& {Weare}(2010)}]{goodmanandweare}
{Goodman}, J., \& {Weare}, J. 2010, Comm. App. Math. Comp. Sci., 5, 65

\bibitem[{{Guidi} {et~al.}(2016){Guidi}, {Tazzari}, {Testi}, {de
  Gregorio-Monsalvo}, {Chandler}, {P{\'e}rez}, {Isella}, {Natta}, {Ortolani},
  {Henning}, {Corder}, {Linz}, {Andrews}, {Wilner}, {Ricci}, {Carpenter},
  {Sargent}, {Mundy}, {Storm}, {Calvet}, {Dullemond}, {Greaves}, {Lazio},
  {Deller}, \& {Kwon}}]{2016A&A...588A.112G}
{Guidi}, G., {Tazzari}, M., {Testi}, L., {et~al.} 2016, \aap, 588, A112

\bibitem[{{Guilloteau} {et~al.}(2011){Guilloteau}, {Dutrey}, {Pi{\'e}tu}, \&
  {Boehler}}]{2011A&A...529A.105G}
{Guilloteau}, S., {Dutrey}, A., {Pi{\'e}tu}, V., \& {Boehler}, Y. 2011, \aap,
  529, A105

\bibitem[{{Hartmann} {et~al.}(1998){Hartmann}, {Calvet}, {Gullbring}, \&
  {D'Alessio}}]{1998ApJ...495..385H}
{Hartmann}, L., {Calvet}, N., {Gullbring}, E., \& {D'Alessio}, P. 1998, \apj,
  495, 385

\bibitem[{{Hu{\'e}lamo} {et~al.}(2010){Hu{\'e}lamo}, {Bouy}, {Pinte},
  {M{\'e}nard}, {Duch{\^e}ne}, {Comer{\'o}n}, {Fern{\'a}ndez}, {Barrado},
  {Bayo}, {de Gregorio-Monsalvo}, \& {Olofsson}}]{2010A&A...523A..42H}
{Hu{\'e}lamo}, N., {Bouy}, H., {Pinte}, C., {et~al.} 2010, \aap, 523, A42

\bibitem[{{Hughes} {et~al.}(2008){Hughes}, {Wilner}, {Qi}, \&
  {Hogerheijde}}]{2008ApJ...678.1119H}
{Hughes}, A.~M., {Wilner}, D.~J., {Qi}, C., \& {Hogerheijde}, M.~R. 2008, \apj,
  678, 1119

\bibitem[{{Isella} {et~al.}(2010){Isella}, {Carpenter}, \&
  {Sargent}}]{2010ApJ...714.1746I}
{Isella}, A., {Carpenter}, J.~M., \& {Sargent}, A.~I. 2010, \apj, 714, 1746

\bibitem[{{Kenyon} \& {Hartmann}(1987)}]{1987ApJ...323..714K}
{Kenyon}, S.~J., \& {Hartmann}, L. 1987, \apj, 323, 714

\bibitem[{{Krist}(1995)}]{1995ASPC...77..349K}
{Krist}, J. 1995, in Astronomical Society of the Pacific Conference Series,
  Vol.~77, Astronomical Data Analysis Software and Systems IV, ed. R.~A.
  {Shaw}, H.~E. {Payne}, \& J.~J.~E. {Hayes}, 349

\bibitem[{{Lebreton} {et~al.}(2012){Lebreton}, {Augereau}, {Thi}, {Roberge},
  {Donaldson}, {Schneider}, {Maddison}, {M{\'e}nard}, {Riviere-Marichalar},
  {Mathews}, {Kamp}, {Pinte}, {Dent}, {Barrado}, {Duch{\^e}ne}, {Gonzalez},
  {Grady}, {Meeus}, {Pantin}, {Williams}, \& {Woitke}}]{2012A&A...539A..17L}
{Lebreton}, J., {Augereau}, J.-C., {Thi}, W.-F., {et~al.} 2012, \aap, 539, A17

\bibitem[{{Long} {et~al.}(2017){Long}, {Herczeg}, {Pascucci}, {Drabek-Maunder},
  {Mohanty}, {Testi}, {Apai}, {Hendler}, {Henning}, {Manara}, \&
  {Mulders}}]{2017arXiv170603320L}
{Long}, F., {Herczeg}, G.~J., {Pascucci}, I., {et~al.} 2017, ArXiv e-prints,
  arXiv:1706.03320

\bibitem[{{Luhman}(2007)}]{2007ApJS..173..104L}
{Luhman}, K.~L. 2007, \apjs, 173, 104

\bibitem[{{Luhman} {et~al.}(2008){Luhman}, {Allen}, {Allen}, {Gutermuth},
  {Hartmann}, {Mamajek}, {Megeath}, {Myers}, \& {Fazio}}]{2008ApJ...675.1375L}
{Luhman}, K.~L., {Allen}, L.~E., {Allen}, P.~R., {et~al.} 2008, \apj, 675, 1375

\bibitem[{{Manara} {et~al.}(2017){Manara}, {Testi}, {Herczeg}, {Pascucci},
  {Alcala}, {Natta}, {Antoniucci}, {Fedele}, {Mulders}, {Henning}, {Mohanty},
  {Prusti}, \& {Rigliaco}}]{2017arXiv170402842M}
{Manara}, C.~F., {Testi}, L., {Herczeg}, G.~J., {et~al.} 2017, ArXiv e-prints,
  arXiv:1704.02842

\bibitem[{{McCabe} {et~al.}(2011){McCabe}, {Duch{\^e}ne}, {Pinte},
  {Stapelfeldt}, {Ghez}, \& {M{\'e}nard}}]{2011ApJ...727...90M}
{McCabe}, C., {Duch{\^e}ne}, G., {Pinte}, C., {et~al.} 2011, \apj, 727, 90

\bibitem[{{Millar-Blanchaer} {et~al.}(2015){Millar-Blanchaer}, {Graham},
  {Pueyo}, {Kalas}, {Dawson}, {Wang}, {Perrin}, {moon}, {Macintosh}, {Ammons},
  {Barman}, {Cardwell}, {Chen}, {Chiang}, {Chilcote}, {Cotten}, {De Rosa},
  {Draper}, {Dunn}, {Duch{\^e}ne}, {Esposito}, {Fitzgerald}, {Follette},
  {Goodsell}, {Greenbaum}, {Hartung}, {Hibon}, {Hinkley}, {Ingraham},
  {Jensen-Clem}, {Konopacky}, {Larkin}, {Long}, {Maire}, {Marchis}, {Marley},
  {Marois}, {Morzinski}, {Nielsen}, {Palmer}, {Oppenheimer}, {Poyneer},
  {Rajan}, {Rantakyr{\"o}}, {Ruffio}, {Sadakuni}, {Saddlemyer}, {Schneider},
  {Sivaramakrishnan}, {Soummer}, {Thomas}, {Vasisht}, {Vega}, {Wallace},
  {Ward-Duong}, {Wiktorowicz}, \& {Wolff}}]{2015ApJ...811...18M}
{Millar-Blanchaer}, M.~A., {Graham}, J.~R., {Pueyo}, L., {et~al.} 2015, \apj,
  811, 18

\bibitem[{{Milli} {et~al.}(2015){Milli}, {Mawet}, {Pinte}, {Lagrange},
  {Mouillet}, {Girard}, {Augereau}, {De Boer}, {Pueyo}, \&
  {Choquet}}]{2015A&A...577A..57M}
{Milli}, J., {Mawet}, D., {Pinte}, C., {et~al.} 2015, \aap, 577, A57

\bibitem[{{Muzerolle} {et~al.}(2009){Muzerolle}, {Flaherty}, {Balog}, {Furlan},
  {Smith}, {Allen}, {Calvet}, {D'Alessio}, {Megeath}, {Muench}, {Rieke}, \&
  {Sherry}}]{2009ApJ...704L..15M}
{Muzerolle}, J., {Flaherty}, K., {Balog}, Z., {et~al.} 2009, \apjl, 704, L15

\bibitem[{{Ossenkopf} \& {Henning}(1994)}]{1994A&A...291..943O}
{Ossenkopf}, V., \& {Henning}, T. 1994, \aap, 291, 943

\bibitem[{{Pascucci} {et~al.}(2016){Pascucci}, {Testi}, {Herczeg}, {Long},
  {Manara}, {Hendler}, {Mulders}, {Krijt}, {Ciesla}, {Henning}, {Mohanty},
  {Drabek-Maunder}, {Apai}, {Sz{\H u}cs}, {Sacco}, \&
  {Olofsson}}]{2016ApJ...831..125P}
{Pascucci}, I., {Testi}, L., {Herczeg}, G.~J., {et~al.} 2016, \apj, 831, 125

\bibitem[{{Pinte} {et~al.}(2009){Pinte}, {Harries}, {Min}, {Watson},
  {Dullemond}, {Woitke}, {M{\'e}nard}, \&
  {Dur{\'a}n-Rojas}}]{2009A&A...498..967P}
{Pinte}, C., {Harries}, T.~J., {Min}, M., {et~al.} 2009, \aap, 498, 967

\bibitem[{{Pinte} {et~al.}(2006){Pinte}, {M{\'e}nard}, {Duch{\^e}ne}, \&
  {Bastien}}]{2006A&A...459..797P}
{Pinte}, C., {M{\'e}nard}, F., {Duch{\^e}ne}, G., \& {Bastien}, P. 2006, \aap,
  459, 797

\bibitem[{{Pinte} {et~al.}(2008){Pinte}, {Padgett}, {M{\'e}nard},
  {Stapelfeldt}, {Schneider}, {Olofsson}, {Pani{\'c}}, {Augereau},
  {Duch{\^e}ne}, {Krist}, {Pontoppidan}, {Perrin}, {Grady}, {Kessler-Silacci},
  {van Dishoeck}, {Lommen}, {Silverstone}, {Hines}, {Wolf}, {Blake}, {Henning},
  \& {Stecklum}}]{2008A&A...489..633P}
{Pinte}, C., {Padgett}, D.~L., {M{\'e}nard}, F., {et~al.} 2008, \aap, 489, 633

\bibitem[{{Pohl} {et~al.}(2017){Pohl}, {Sissa}, {Langlois}, {M{\"u}ller},
  {Ginski}, {van Holstein}, {Vigan}, {Mesa}, {Maire}, {Henning}, {Gratton},
  {Olofsson}, {van Boekel}, {Benisty}, {Biller}, {Boccaletti}, {Chauvin},
  {Daemgen}, {de Boer}, {Desidera}, {Dominik}, {Garufi}, {Janson}, {Kral},
  {M{\'e}nard}, {Pinte}, {Stolker}, {Szul{\'a}gyi}, {Zurlo}, {Bonnefoy},
  {Cheetham}, {Cudel}, {Feldt}, {Kasper}, {Lagrange}, {Perrot}, \&
  {Wildi}}]{2017A&A...605A..34P}
{Pohl}, A., {Sissa}, E., {Langlois}, M., {et~al.} 2017, \aap, 605, A34

\bibitem[{{Raftery} \& {Lewis}(1992)}]{bayesianstatistics}
{Raftery}, A., \& {Lewis}, S. 1992, Bayesian Statistics, 4, 763

\bibitem[{{Ribas} {et~al.}(2016){Ribas}, {Bouy}, {Mer{\'{\i}}n}, {Duch{\^e}ne},
  {Rebollido}, {Espaillat}, \& {Pinte}}]{2016MNRAS.458.1029R}
{Ribas}, {\'A}., {Bouy}, H., {Mer{\'{\i}}n}, B., {et~al.} 2016, \mnras, 458,
  1029

\bibitem[{{Ricci} {et~al.}(2015){Ricci}, {Carpenter}, {Fu}, {Hughes}, {Corder},
  \& {Isella}}]{2015ApJ...798..124R}
{Ricci}, L., {Carpenter}, J.~M., {Fu}, B., {et~al.} 2015, \apj, 798, 124

\bibitem[{{Robberto} {et~al.}(2012){Robberto}, {Spina}, {Da Rio}, {Apai},
  {Pascucci}, {Ricci}, {Goddi}, {Testi}, {Palla}, \&
  {Bacciotti}}]{2012AJ....144...83R}
{Robberto}, M., {Spina}, L., {Da Rio}, N., {et~al.} 2012, \aj, 144, 83

\bibitem[{{Robitaille} {et~al.}(2006){Robitaille}, {Whitney}, {Indebetouw},
  {Wood}, \& {Denzmore}}]{2006ApJS..167..256R}
{Robitaille}, T.~P., {Whitney}, B.~A., {Indebetouw}, R., {Wood}, K., \&
  {Denzmore}, P. 2006, \apjs, 167, 256

\bibitem[{{Rodgers-Lee} {et~al.}(2014){Rodgers-Lee}, {Scholz}, {Natta}, \&
  {Ray}}]{2014MNRAS.443.1587R}
{Rodgers-Lee}, D., {Scholz}, A., {Natta}, A., \& {Ray}, T. 2014, \mnras, 443,
  1587

\bibitem[{{Sauter} {et~al.}(2009){Sauter}, {Wolf}, {Launhardt}, {Padgett},
  {Stapelfeldt}, {Pinte}, {Duch{\^e}ne}, {M{\'e}nard}, {McCabe}, {Pontoppidan},
  {Dunham}, {Bourke}, \& {Chen}}]{2009A&A...505.1167S}
{Sauter}, J., {Wolf}, S., {Launhardt}, R., {et~al.} 2009, \aap, 505, 1167

\bibitem[{{Sharma}(2017)}]{2017arXiv170601629S}
{Sharma}, S. 2017, ArXiv e-prints, arXiv:1706.01629

\bibitem[{{Stapelfeldt}(2004)}]{2004IAUS..202..291S}
{Stapelfeldt}, K. 2004, in IAU Symposium, Vol. 202, Planetary Systems in the
  Universe, ed. A.~{Penny}, 291

\bibitem[{{Stapelfeldt} {et~al.}(2014){Stapelfeldt}, {Duch{\^e}ne}, {Perrin},
  {Wolff}, {Krist}, {Padgett}, {M{\'e}nard}, \& {Pinte}}]{2014IAUS..299...99S}
{Stapelfeldt}, K.~R., {Duch{\^e}ne}, G., {Perrin}, M., {et~al.} 2014, in IAU
  Symposium, Vol. 299, Exploring the Formation and Evolution of Planetary
  Systems, ed. M.~{Booth}, B.~C. {Matthews}, \& J.~R. {Graham}, 99--103

\bibitem[{{Steinacker} {et~al.}(2013){Steinacker}, {Baes}, \&
  {Gordon}}]{2013ARA&A..51...63S}
{Steinacker}, J., {Baes}, M., \& {Gordon}, K.~D. 2013, \araa, 51, 63

\bibitem[{{van der Plas} {et~al.}(2016){van der Plas}, {M{\'e}nard},
  {Ward-Duong}, {Bulger}, {Harvey}, {Pinte}, {Patience}, {Hales}, \&
  {Casassus}}]{2016ApJ...819..102V}
{van der Plas}, G., {M{\'e}nard}, F., {Ward-Duong}, K., {et~al.} 2016, \apj,
  819, 102

\bibitem[{{Watson} {et~al.}(2007){Watson}, {Stapelfeldt}, {Wood}, \&
  {M{\'e}nard}}]{2007prpl.conf..523W}
{Watson}, A.~M., {Stapelfeldt}, K.~R., {Wood}, K., \& {M{\'e}nard}, F. 2007,
  Protostars and Planets V, 523

\bibitem[{{Weidenschilling}(1977)}]{1977Ap&SS..51..153W}
{Weidenschilling}, S.~J. 1977, \apss, 51, 153

\bibitem[{{Whitney} {et~al.}(2004){Whitney}, {Indebetouw}, {Bjorkman}, \&
  {Wood}}]{2004ApJ...617.1177W}
{Whitney}, B.~A., {Indebetouw}, R., {Bjorkman}, J.~E., \& {Wood}, K. 2004,
  \apj, 617, 1177

\bibitem[{{Whitney} {et~al.}(2003{\natexlab{a}}){Whitney}, {Wood}, {Bjorkman},
  \& {Cohen}}]{2003ApJ...598.1079W}
{Whitney}, B.~A., {Wood}, K., {Bjorkman}, J.~E., \& {Cohen}, M.
  2003{\natexlab{a}}, \apj, 598, 1079

\bibitem[{{Whitney} {et~al.}(2003{\natexlab{b}}){Whitney}, {Wood}, {Bjorkman},
  \& {Wolff}}]{2003ApJ...591.1049W}
{Whitney}, B.~A., {Wood}, K., {Bjorkman}, J.~E., \& {Wolff}, M.~J.
  2003{\natexlab{b}}, \apj, 591, 1049

\bibitem[{{Whittet} {et~al.}(1997){Whittet}, {Prusti}, {Franco}, {Gerakines},
  {Kilkenny}, {Larson}, \& {Wesselius}}]{1997A&A...327.1194W}
{Whittet}, D.~C.~B., {Prusti}, T., {Franco}, G.~A.~P., {et~al.} 1997, \aap,
  327, 1194

\bibitem[{{Williams} \& {Cieza}(2011)}]{2011ARA&A..49...67W}
{Williams}, J.~P., \& {Cieza}, L.~A. 2011, \araa, 49, 67

\bibitem[{{Winston} {et~al.}(2012){Winston}, {Cox}, {Prusti}, {Mer{\'{\i}}n},
  {Ribas}, {Royer}, {Vavrek}, {Puga}, {Andr{\'e}}, {Men'shchikov},
  {K{\"o}nyves}, {K{\'o}sp{\'a}l}, {Alves de Oliveira}, {Pilbratt}, \&
  {Waelkens}}]{2012A&A...545A.145W}
{Winston}, E., {Cox}, N.~L.~J., {Prusti}, T., {et~al.} 2012, \aap, 545, A145

\bibitem[{{Woitke}(2015)}]{2015EPJWC.10200007W}
{Woitke}, P. 2015, in European Physical Journal Web of Conferences, Vol. 102,
  European Physical Journal Web of Conferences, 00007

\bibitem[{{Woitke} {et~al.}(2010){Woitke}, {Pinte}, {Tilling}, {M{\'e}nard},
  {Kamp}, {Thi}, {Duch{\^e}ne}, \& {Augereau}}]{2010MNRAS.405L..26W}
{Woitke}, P., {Pinte}, C., {Tilling}, I., {et~al.} 2010, \mnras, 405, L26

\bibitem[{{Wolf} {et~al.}(2003){Wolf}, {Padgett}, \&
  {Stapelfeldt}}]{2003ApJ...588..373W}
{Wolf}, S., {Padgett}, D.~L., \& {Stapelfeldt}, K.~R. 2003, \apj, 588, 373

\bibitem[{{Wolff} {et~al.}(2017){Wolff}, {Perrin}, {Ren}, \&
  {Pinte}}]{mcfost-python}
{Wolff}, S.~G., {Perrin}, M., {Ren}, B., \& {Pinte}, C. 2017, {mcfost-python},
  v1.0,  Zenodo, doi:10.5281/zenodo.839863.
\newblock \url{https://doi.org/10.5281/zenodo.839863}

\bibitem[{{Wood} {et~al.}(2008){Wood}, {Whitney}, {Robitaille}, \&
  {Draine}}]{2008ApJ...688.1118W}
{Wood}, K., {Whitney}, B.~A., {Robitaille}, T., \& {Draine}, B.~T. 2008, \apj,
  688, 1118

\end{thebibliography}

\end{document}